\documentclass[12pt]{article}
\usepackage{amsfonts}
\usepackage{amsmath, amsthm}
\usepackage{epsfig}
\usepackage{algorithm}
\usepackage{algorithmic}
\usepackage{epic}
\usepackage[a4paper, bottom = 1.50cm,
        ]{geometry}
\usepackage{enumitem}

\usepackage{amsmath,verbatim,color,amssymb,epsfig}
\usepackage{bm}
\usepackage{ragged2e}
\usepackage{amsfonts}
\usepackage{epsfig}
\usepackage{changepage}
\usepackage{multirow}
\usepackage{graphicx}
\usepackage{array}
\usepackage[table]{xcolor}
\definecolor{Gray}{gray}{.80}
\usepackage{lscape}
\usepackage[round,semicolon,authoryear]{natbib}
\usepackage[normalem]{ulem}
\usepackage[colorlinks=true,urlcolor=blue,citecolor=purple,linkcolor=blue,bookmarks=true]{hyperref}
\usepackage{threeparttable}
\usepackage{pdfsync}
\usepackage{array}

\usepackage{booktabs}  
\usepackage{cleveref} 
\usepackage{caption}  
\usepackage{tabularx}   
\usepackage{adjustbox} 
\usepackage{longtable}  

\begin{document}
\def\eqx"#1"{{\label{#1}}}
\def\eqn"#1"{{\ref{#1}}}

\makeatletter 
\@addtoreset{equation}{section}
\makeatother  

\def\yuancomment#1{\vskip 2mm\boxit{\vskip 2mm{\color{red}\bf#1} {\color{blue}\bf --Yuan\vskip 2mm}}\vskip 2mm}
\def\lincomment#1{\vskip 2mm\boxit{\vskip 2mm{\color{blue}\bf#1} {\color{black}\bf --Lin\vskip 2mm}}\vskip 2mm}
\def\squarebox#1{\hbox to #1{\hfill\vbox to #1{\vfill}}}
\def\boxit#1{\vbox{\hrule\hbox{\vrule\kern6pt
          \vbox{\kern6pt#1\kern6pt}\kern6pt\vrule}\hrule}}

\def\theequation{\thesection.\arabic{equation}}
\newcommand{\ds}{\displaystyle}

\newcommand{\bJ}{\mbox{\bf J}}
\newcommand{\bF}{\mbox{\bf F}}
\newcommand{\bM}{\mbox{\bf M}}
\newcommand{\bR}{\mbox{\bf R}}
\newcommand{\bZ}{\mbox{\bf Z}}
\newcommand{\bX}{\mbox{\bf X}}
\newcommand{\bx}{\mbox{\bf x}}
\newcommand{\bww}{\mbox{\bf w}}
\newcommand{\bQ}{\mbox{\bf Q}}
\newcommand{\bH}{\mbox{\bf H}}
\newcommand{\bh}{\mbox{\bf h}}
\newcommand{\bz}{\mbox{\bf z}}
\newcommand{\br}{\mbox{\bf r}}
\newcommand{\ba}{\mbox{\bf a}}
\newcommand{\be}{\mbox{\bf e}}
\newcommand{\bG}{\mbox{\bf G}}
\newcommand{\bB}{\mbox{\bf B}}
\newcommand{\bb}{\mbox{\bf b}}
\newcommand{\bA}{\mbox{\bf A}}
\newcommand{\bC}{\mbox{\bf C}}
\newcommand{\bI}{\mbox{\bf I}}
\newcommand{\bD}{\mbox{\bf D}}
\newcommand{\bU}{\mbox{\bf U}}
\newcommand{\bc}{\mbox{\bf c}}
\newcommand{\bd}{\mbox{\bf d}}
\newcommand{\bs}{\mbox{\bf s}}
\newcommand{\bS}{\mbox{\bf S}}
\newcommand{\bV}{\mbox{\bf V}}
\newcommand{\bv}{\mbox{\bf v}}
\newcommand{\bW}{\mbox{\bf W}}
\newcommand{\bY}{\mathbf{ Y}}
\newcommand{\bw}{\mbox{\bf w}}
\newcommand{\bg}{\mbox{\bf g}}
\newcommand{\bu}{\mbox{\bf u}}
\newcommand{\mI}{\mbox{I}}

\def\bb{{\bf b}}

\newcommand{\bcU}{\boldsymbol{\cal U}}
\newcommand{\bbeta}{\boldsymbol{\beta}}
\newcommand{\bdelta}{\boldsymbol{\Delta}}
\newcommand{\bDelta}{\boldsymbol{\Delta}}
\newcommand{\boldeta}{\boldsymbol{\eta}}
\newcommand{\bxi}{\boldsymbol{\xi}}
\newcommand{\bGamma}{\boldsymbol{\Gamma}}
\newcommand{\bSigma}{\boldsymbol{\Sigma}}
\newcommand{\balpha}{\boldsymbol{\alpha}}
\newcommand{\bOmega}{\boldsymbol{\Omega}}
\newcommand{\btheta}{\boldsymbol{\theta}}
\newcommand{\bepsilon}{\boldsymbol{\epsilon}}
\newcommand{\bmu}{\boldsymbol{\mu}}
\newcommand{\bnu}{\boldsymbol{\nu}}
\newcommand{\bgamma}{\boldsymbol{\gamma}}
\newcommand{\btau}{\boldsymbol{\tau}}
\newcommand{\bTheta}{\boldsymbol{\Theta}}

\newtheorem{thm}{Theorem}
\newtheorem{lem}{Lemma}[section]
\newtheorem{rem}{Remark}[section]
\newtheorem{cor}{Corollary}[section]
\newcolumntype{L}[1]{>{\raggedright\let\newline\\\arraybackslash\hspace{0pt}}m{#1}}
\newcolumntype{C}[1]{>{\centering\let\newline\\\arraybackslash\hspace{0pt}}m{#1}}
\newcolumntype{R}[1]{>{\raggedleft\let\newline\\\arraybackslash\hspace{0pt}}m{#1}}

\newcommand{\tabincell}[2]{\begin{tabular}{@{}#1@{}}#2\end{tabular}}

\newcommand{\RN}[1]{%
  \textup{\uppercase\expandafter{\romannumeral#1}}%
}

\newcommand{\lline}[1]{\hline\multicolumn{#1}{c}{}\\[-1.3em]\hline}

\baselineskip=14pt
\begin{center}
{\Large \bf BOP2-TE: Bayesian Optimal Phase 2 Design for Jointly Monitoring Efficacy and Toxicity with Application to Dose Optimization}
\end{center}

\vspace{1mm}
\begin{center}
{\bf Kai Chen$^{1,3}$, Heng Zhou$^{2}$, J. Jack Lee$^{3}$, Ying Yuan$^{3,*}$}
\end{center}

\noindent$^{1}$Department of Biostatistics and Data Science, The University of Texas Health Science center, Houston, TX\\
$^{2}$Biostatistics and Research Decision Sciences, Merck $\&$ Co., Inc, Kenilworth, New Jersey \\
$^{3}$Department of Biostatistics, The University of Texas MD Anderson Cancer Center, Houston, TX \\
$^*$Author for correspondence: yyuan@mdanderson.org 
\vspace{2mm}

\baselineskip=24pt

\noindent \emph{\textbf{Abstract}}:  We propose a Bayesian optimal phase 2 design for jointly monitoring efficacy and toxicity, referred to as BOP2-TE, to improve the operating characteristics of the BOP2 design proposed by Zhou et al. (2017). BOP2-TE utilizes a Dirichlet-multinomial model to jointly model the distribution of toxicity and efficacy endpoints, making go/no-go decisions based on the posterior probability of toxicity and futility. In comparison to the original BOP2 and other existing designs, BOP2-TE offers the advantage of providing rigorous type I error control in cases where the treatment is toxic and futile, effective but toxic, or safe but futile, while optimizing power when the treatment is effective and safe. As a result, BOP2-TE enhances trial safety and efficacy. We also explore the incorporation of BOP2-TE into multiple-dose randomized trials for dose optimization, and consider a seamless design that integrates phase I dose finding with phase II randomized dose optimization. BOP2-TE is user-friendly, as its decision boundary can be determined prior to the trial's onset. Simulations demonstrate that BOP2-TE possesses desirable operating characteristics. We have developed a user-friendly web application as part of the BOP2 app, which is freely available at www.trialdesign.org.\\


\section{Introduction}
\label{introduction}

The primary objective of phase II trials is to evaluate the preliminary evidence of efficacy of a new drug, which is administrated at the maximum tolerated dose (MTD) or recommended phase II dose (RP2D), established in phase I trials. The purpose is to determine whether the drug shows sufficient promise for further study in phase III study. To simplify this go/no-go decision and shorten the trial duration, most phase II trial designs, such as Simon's optimal two-stage design \citep{Simon1989}, focus on a single, binary short-term efficacy endpoints (e.g., response or non-response). The go/no-go decision is based on whether the response rate reaches a specific desirable level, without explicitly incorporating safety into the decision making. 

Although the MTD/RP2D evaluated in phase II trials has gone through initial safety assessment in phase I trials, due to small sample size, there is still a high degree of uncertainty regarding its safety. For instance, if 1 out of 6 patients experienced dose-limiting toxicity at the MTD/RP2D, the 95\% exact confidence interval for the toxicity rate of the dose would be (0.004, 0.641). To ensure patient safety and maximize the treatment benefit, it is crucial to incorporate safety considerations, in addition to efficacy, into the go/no-go decisions. Recognizing this, the U.S. Food and Drug Administration (FDA) has issued guidance on ``Benefit-Risk Assessment for New Drug and Biological Products" to underscore the importance of considering both efficacy and safety in making go/no-go decisions in drug development.

A number of designs have been proposed to formally incorporate toxicity into phase II trials by jointly monitoring toxicity and efficacy. \citet{Bryant1993} developed a frequentist two-stage sequential design to jointly evaluate binary efficacy and toxicity endpoints, which minimizes the expected sample size under the null hypothesis. 
\citet{Conaway1995} proposed a toxicity-efficacy bivariate sequential design that accommodates more than two stages. \citet{Ray2012} combined Simon's two-stage design for efficacy endpoint and a continuous toxicity monitoring method for toxicity endpoint. \citet{Thall1995} introduced a Bayesian sequential monitoring design to joint model efficacy and toxicity endpoints based on Dirichlet-multinomial model. \citet{Zhou2017} and \citet{Zhou2020} developed Bayesian optimal phase II design (BOP2) as a unified approach to handle various types of endpoints (e.g., binary, ordinal, survival, co-primary, and multiple endpoints), including jointly monitoring toxicity and efficacy.  \citet{Zhao2023} extended BOP2 to randomized phase II trials and enable go/consider/no-go decisions based on dual criteria that consider both statistical and clinical significance. BOP2 design is efficient and highly flexible. It allows any arbitrary number of interim looks and maximizes power with controlled type I error. To the best of our knowledge, over 50 phase II or I-II trials in the MD Anderson Cancer Center are based on BOP2.

 One limitation of using the BOP2 design to jointly monitor toxicity and efficacy is that it only controls type I error under the ``global null," where the drug is both futile and toxic. It does not provide explicit control over type I error under ``partial nulls", where the drug is either safe but futile or efficacious but toxic. As a result, in certain scenarios described later, although the probability of a go decision is controlled at the target level of 5\% under the global null, the probability of a go decision can be as high as 20\% under the partial null that the drug is efficacious but toxic. 

To address this issue, this paper develops an extension of BOP2, called BOP2-TE, that enables investigators to control type I error under the global null and partial nulls. Although we focus on binary toxicity and efficacy endpoints, the design can be extended to multinomial endpoints as discussed in the discussion section. With BOP2-TE, investigators specify three target type I error rates corresponding to the global null and two partial nulls, respectively. BOP2-TE makes go/no-go decisions at interim and the end of the trial based on Bayesian posterior probabilities that the drug is safe and efficacious. The stopping boundaries are constructed to maximize power while controlling type I error at desirable levels under both global null and partial nulls. 

While BOP2-TE employs a similar statistical model and posterior probabilities to BOP2 to make go/no-go decisions, this paper provides several new contributions. Firstly, BOP2-TE better reflects clinical practice, thereby enhancing patient safety and benefit. To facilitate the design's application, we have developed freely available software at www.trialdesign.org. Secondly, we investigate the relationship between type I error and the correlation between toxicity and efficacy, and our findings provide useful guidance on using both BOP2-TE and BOP2. Lastly, we derive closed-form type I error, which allows for rapid and accurate optimization of stopping boundaries. In contrast, the original BOP2 paper relies on Monte Carlo simulation to calculate type I error, which is computationally intensive. 

Although we focus on phase II trials, BOP2-TE can be used to monitor cohort expansion, which is increasingly popular in dose optimization trials. We demonstrate this application in  randomized  dose optimization trials, where patients are randomized into multiple doses for identifying the optimal dose. This approach has been recommended by FDA's guidance on dose optimization \citep{FDAoptimus}. We also briefly investigate a seamless dose optimization design that integrates phase I dose finding with phase II randomized dose optimization using the BOP2-TE.

In Section 2, we describe the Bayesian model for toxicity and efficacy, the algorithm for parameter optimization, and software for implementing the resulting the BOP2-TE design.  In Section 3, we discuss BOP2-TE for dose optimization.  In Section 4, we present a simulation study that evaluates the operating characteristics of BOP2-TE for toxicity and efficacy monitoring, as well as dose optimization. We conclude with a discussion in Section 5.  

\section{Method}\label{method}
\subsection{Efficacy-toxicity model} \label{Efftoxmodel}


Let $Y_E$ denote a binary efficacy endpoint with $Y_E =1$ indicating response, and $Y_T$ denote a binary toxicity endpoint with $Y_T=1$ indicating toxicity. The bivariate outcomes $(Y_E,Y_T)$ can be represented by a multinomial variable $Y = (y_1,...,y_K)$ with $K=4$ possible outcomes with $y_1=1$ if $(Y_E,Y_T)=(1,1)$; $y_2=1$ if $(Y_E,Y_T)=(1,0)$; $y_3=1$ if $(Y_E,Y_T)=(0,1)$; $y_4=1$ if $(Y_E,Y_T)=(0,0)$, as follows:
$$
    Y \sim \text{Multinomial } (\pi_1,...,\pi_K), 
$$
where $\pi_k$ is the probability of $y_k=1$ and other $y_k$'s equal to 0, with $\sum_{k=1}^K \pi_k=1$.  

Suppose that at an interim time, $n$ patients have been treated and $x_k$ patients had outcome $y_k=1$, and let $D_n = \{x_1, ...,x_K\}$ denote the interim data. We assign $(\pi_1,...,\pi_K)$ a Dirichlet prior as 
$$
    \pi_1, ..., \pi_K \sim \text{Dirichlet}(\tau_1, ...,\tau_K),
$$
where $\tau_1,...,\tau_K$ are hyperparameters. The posterior distribution of $(\pi_1,...,\pi_K)$ arises as
$$
\pi_1,...,\pi_K|D_n \sim \text{Dirichlet}(\tau_1+x_1, ...,\tau_k+x_k).
$$
Under the Dirichlet-multinomial model,  $\sum_{k=1}^K \tau_k$ can be interpreted as the prior effective sample size, which facilitate the specification of hyperparameter $\tau_k$. To obtain a vague prior, we set  $\tau_k$ at the prior estimate of $\pi_k$ with $\sum_{k=1}^K \tau_k = 1$ such that the prior information is equivalent to one patient. If prior information is available, an informative prior with a proper effective sample size could be used. 

An important property of the Dirichlet distribution is that any combination of $\pi_k$'s follows a beta distribution. Specifically, define the marginal response rate $\pi_E=\pi_1 + \pi_2$ and the marginal toxicity rate $\pi_T=\pi_1 + \pi_3$. Let $x_E = x_1 + x_2$ denote the number of responses, and $x_T = x_1 + x_3$ denote the number of toxicities. It follows that 
\begin{eqnarray}
\pi_E | D_n \sim \text{Beta}(\tau_E  +x_E, n+1 -\tau_E - x_E)  \label{eq: efficacy}\\
\pi_T | D_n \sim \text{Beta}(\tau_T  +x_T, n+1 -\tau_T - x_T),  \label{eq: toxicity}
\end{eqnarray}
where  $\tau_E = \tau_1 + \tau_2$ and $\tau_T = \tau_1 + \tau_3$. As described next, our hypotheses and go/no-go decisions are based on the posterior distribution of $\pi_E$ and $\pi_T$.

\subsection{Type I error and power} \label{s:type1}
In considering both $Y_E$ and $Y_T$, controlling type I error and power requires additional considerations beyond the univariate case where only $Y_E$ is considered, such as in Simon's optimal two-stage design. 

We first briefly review how the original BOP2 defines and controls type I error based on $Y_E$ and $Y_T$. Let $\eta_{E}$ and $\eta_E^*$ denote the target efficacy rate and unacceptable efficacy rate, respectively, and let $\eta_T$ and $\eta_{T}^*$ denote the desirable toxicity rate and unacceptable toxicity rate, respectively, with $0 < \eta_E^* < \eta_{E}$ and $0 < \eta_T < \eta_{T}^*$. BOP2 considers the following null and alternative hypotheses,
\begin{eqnarray}
H_{00}: & \pi_E = \eta_{E}^*, \,\, \pi_T = \eta_T^* \label{H00} \\
H_{11}: & \pi_E = \eta_{E}, \,\, \pi_T = \eta_T, \label{H11}
\end{eqnarray}
where $H_{00}$ represents the global null that the treatment is futile and toxic, and $H_{11}$ represents the alternative that the treatment is safe and efficacy. BOP2 defines type I error and power respectively as 
$$\alpha_{00} = Pr(\text{claim promising} \,|\, H_{00})$$ 
$$\beta = Pr(\text{claim promising} \,|\, H_{11}),$$  
where ``claim promising" denote claiming that the treatment is promising (i.e., efficacious and safe) at the end of the trial.  BOP2 identifies the optimal go/no-go boundaries that control $\alpha_{00} \le \alpha_{00}^*$ and maximize power, where $\alpha_{00}^*$ is a prespecified level (e.g., 5\% or 10\%).  

Unfortunately, $H_{00}$ alone  is not sufficient to cover all null scenarios that a no-go decision should be made, and thus controlling $\alpha_{00}$ only does not fully cover the ethical consideration for futility and toxicity.  To illustrate this point, consider a phase II trial where the null and target response rates are $0.3$ and $0.6$, respectively, and the null and acceptable toxicity rate is $0.4$ and $0.2$, respectively. The maximum sample size is 36, with an interim analysis when 18 subjects are enrolled. To control the type I error rate at 5\% under $H_{00}$ of $\pi_E=0.3$ and $\pi_T=0.4$, BOP2 yields the optimal go/no-go boundaries: requiring the number of response greater than 4 in interim time and 12 in the end, while keeping the number of toxicity less than 7 and 13, respectively. However, these go/no-go boundaries lead to a go decision $19\%$ of time when $\pi_E=0.3$ and $\pi_T=0.2$ (i.e., futile and safe), and $21\%$ of time when $\pi_E=0.6$ and $\pi_T=0.4$ (i.e., efficacious but unacceptably toxic). 

To address this issue, in BOP2-TE, we introduce two additional partial null hypotheses:
\begin{eqnarray}
 H_{01}: & \pi_E = \eta_E^*, \,\, \pi_{T} =\eta_T\\
 H_{10}: & \pi_E = \eta_{E}, \,\, \pi_{T} = \eta_{T}^*,
\end{eqnarray}
where $H_{01}$ represents that the treatment is safe but futile, and $H_{10}$ represents that the treatment is efficacious but toxic. Accordingly, we define two additional type I errors: 
$$\alpha_{01} = Pr(\text{claim promising} \,|\, H_{01})$$
$$\alpha_{10} = Pr(\text{claim promising} \,|\, H_{10}).$$ 
To ensure ethical considerations for futility and toxicity are fully covered, the BOP2-TE controls all three type I errors:
\begin{eqnarray}
\alpha_{00}  &\le& \alpha_{00}^* \notag \\
\alpha_{01} &\le& \alpha_{01}^*     \label{eq:constraint}\\
\alpha_{10} &\le& \alpha_{10}^*, \notag
\end{eqnarray}
where $\alpha_{01}^*$ and $\alpha_{10}^*$ are desirable levels for $\alpha_{01}$ and $\alpha_{10}$, respectively.  These values should be chosen based on the disease and target patient population to reflect the risk-benefit assessment underlying clinical decisions. In general, we should set $\alpha_{01}^* \ge \alpha_{00}^*$ and $\alpha_{10}^* \ge \alpha_{00}^*$ because $H_{01}$ and $H_{10}$ are less of concern as $H_{00}$. For example, when $\alpha_{00}^*=0.05$, we may set $\alpha_{01}^*=0.10$ and $\alpha_{10}^*=0.10$.

Controlling type I errors under both global and partial null hypotheses not only provides better protection for patients, but also offers investigators additional flexibility to customize the trial design to account for risk-benefit tradeoff of the treament by selecting appropriate target type I error levels. For instance, in a therapeutic area where there is a lack of effective treatments and patients have a high tolerance for potentially futile treatments, a larger $\alpha^*_{01}$ might be used, such as $\alpha_{00}^*=0.05$, $\alpha_{01}^*=0.20$, and $\alpha_{10}^*=0.10$. Conversely, if the toxicity of the treatment is generally not severe and manageable, a higher $\alpha^*_{10}$ might be considered, like $\alpha_{00}^*=0.05$, $\alpha_{01}^*=0.10$, and $\alpha_{10}^*=0.20$. 

The specification of $\alpha_{00}^*$, $\alpha_{01}^*$, and $\alpha_{10}^*$ is facilitated by their intuitive interpretations, which represent the tolerable false-go rates when the treatment is toxic and futile,  safe but futile, and efficacious but toxic, respectively. After eliciting $\alpha_{00}^*$, $\alpha_{01}^*$, and $\alpha_{10}^*$ from clinicians, if needed, their values could be further calibrated using simulations to ensure the desired operating characteristics. The calibration is simplified by the fast-computing software described later in Section \ref{sec:soft}. 

The values of $(\eta_E, \eta_E^*, \eta_T, \eta_T^*)$ control the power of the design. Given the small to modest sample sizes of typical phase II trials, we should refrain from setting the values of $\eta_E$ and $\eta_E^*$, or $\eta_T$ and $\eta_T^*$, too closely to each other, due to limited power to distinguish a small effect size. For instance, given a sample size of 40, there is merely a 16\% power to differentiate $\eta_E= 0.40$ from $\eta_E^*=0.35$ based on a one-sided proportion test with a significance level of 0.05. 

We recommend the following procedure to specify $(\eta_E, \eta_E^*, \eta_T, \eta_T^*)$. We first elicit the lowest acceptable efficacy rate $\phi_E$ and the highest acceptable toxicity rate $\phi_T$ from clinicians, and then set $(\eta_E, \eta_E^*)= (\phi_E+\delta_{E1}, \phi_E-\delta_{E2})$ and $( \eta_T, \eta_T^*)= (\phi_T-\delta_{T1}, \phi_T+\delta_{T2})$, where $\delta_{E1}$, $\delta_{E2}$, $\delta_{T1}$, and $\delta_{T2}$ are margins that reasonably distance $(\eta_E, \eta_E^*, \eta_T, \eta_T^*)$ from $\phi_E$ and $\phi_T$. In many applications,  $\delta_{E1}=\delta_{E2} = 0.15$ or 0.1 and $\delta_{T1}=\delta_{T2} = 0.1$ provide a reasonable default choice,  which can be further calibrated based on the available sample size, target power, and effect size of clinical interest.

Although three target type I errors are specified, it is often unlikely to control $\alpha_{00}$, $\alpha_{01}$, and $\alpha_{10}$ exactly at their target values. In many cases, when two type I errors (e.g.,  $\alpha_{01}$ and $\alpha_{10}$) are controlled at their target values, the remaining one (e.g., $\alpha_{00}$) will be well below its value, automatically satisfying its type I error constraint. One setting of particular interest is when $Y_T$ and $Y_E$ are independent, which we later recommend for optimizing design parameters in Section \ref{sec:optpara}. In this case, it can be shown that  $\alpha_{00}$, $\alpha_{01}$, and $\alpha_{10}$ have the following relationship (see Appendix A.1 for proof): 

\bigskip
\noindent\textbf{Theorem 1.} When $Y_T$ and $Y_E$ are independent, $\alpha_{00} = \frac{\displaystyle \alpha_{01}\alpha_{10}}{\displaystyle  \beta}$, where $\beta$ denotes power.
\bigskip

 As a result, it is quite common that when two of the type I error constraints are satisfied, the remaining one is automatically and conservatively satisfied. For example, when $(\alpha_{00}^*$, $\alpha_{01}^*$, $\alpha_{10}^*$)  = (0.025, 0.10, 0.10), if $\alpha_{01} = \alpha_{01}^*=0.1$ and $\alpha_{10} = \alpha_{10}^*=0.1$, and assuming power is 0.8, then $\alpha_{00} =  0.1\times0.1/0.8=0.012$, which is conservatively lower than $\alpha_{00}^*$. Our simulation later verified this phenomenon.

%
%

\subsection{Trial design}

BOP2-TE is a sequential design with multiple stages. Assume there are $R$ stages, consisting of $R-1$ interim analyses,  conducted when the number of patients reaches $n_1,n_2,..., n_{R-1}$, and a final analysis when the sample size reaches $N$. The go/no-go decision at the interim and final analyses is made based on the following Bayesian criteria :
\begin{quote}
Make go decision if  $\Pr(\pi_E> \eta_E^*  |D_n) > C_E(n)$ and $\Pr(\pi_T \le \eta_T^*  |D_n) > C_T(n)$, otherwise stop the trial and claim that the treatment is not promising,  where $C_E(n)$ and $C_T(n)$ are probabilities cutoffs that are functions of the interim sample size $n$. 
\end{quote}
Here, ``go" means continuing to the next interim before the completion of the trial, and claiming that the treatment is promising at the end of the trial.

Following \citet{Zhou2017}, we employ the following information-dependent decision boundaries:
\begin{eqnarray}
C_E(n) &=& \lambda_E(\frac{n}{N})^{\gamma}  \label{CE}  \\
C_T(n) &=& \lambda_T(\frac{n}{N})^{\frac{\gamma}{3}},  \label{CT}
\end{eqnarray}
where $0\le \lambda_E$, $\lambda_T$, $\gamma \le 1$ are design parameters, whose determination will be discussed in the next section. $\gamma$ controls how the stopping boundary changes with the fraction of information. We impose the constraint $0 \le \gamma \le 1$ to ensure that the boundaries for go decisions at interims are less stringent than at the final analysis but not excessively loose. Specifically, when $\gamma = 0$, we obtain a constant cutoff for interims and final analysis, similar to Pocock boundaries. When $\gamma = 1/2$, the boundaries are proportional to the square root of the fraction of information, similar to O'Brien-Fleming boundaries. In this case, the boundary for go decision at interims is looser than at the final analysis. When $\gamma > 1/2$, the trial is less likely to stop at the interims than with O'Brien-Fleming boundaries, meaning the boundary for go decision at interims is dramatically looser than at the final analysis. In practice, it is very difficult to reach O'Brien-Fleming boundaries for stopping, so it makes little practical sense to have boundaries dramatically looser than that, as it defeats the purpose of interim monitoring. Therefore, we require $\gamma \le 1$. This restriction is used in the original BOP2 design and its software but was not explicitly defined and clearly explained. 

To enhance safety, we added an attenuation factor of 3 to the toxicity probability cutoff (\ref{CT}) to impose stricter stopping boundaries for early interims with small $n$. This approach has been adopted by BOP2 software and used in many ongoing trials based on BOP2, before the development of BOP2-TE here. The selection of an attenuation factor of 3 is based on the following consideration: when $(\eta_T, \eta_T^*)=(0.2, 0.4)$, which suggests that the highest acceptable toxicity rate is about 0.3, the above Bayesian rule generates a safety stopping boundary aligning with the conventional rule that 2/3 or 3/6 toxicities are deemed unacceptable (see Appendix A.2). Setting the attenuation factor of 3 is not required by our methodology; if necessary, it can be adjusted. 

A more flexible approach is to use distinct power parameter $\gamma$ in (\ref{CE}) and (\ref{CT}). However, we opt for the approach of sharing the same $\gamma$ between $C_E(n)$ and $C_T(n)$ for two reasons. First, using a distinct $\gamma$ for $C_T(n)$ often results in overly relaxed toxicity boundaries. For example, when $(\eta_T, \eta_T^*)=(0.2, 0.4)$, the stopping boundary would be 3/3 toxicities. This happens because the algorithm optimizes parameters in $C_T(n)$ (and also $C_E(n)$) to maximize power, as described in the next section. To increase power, the optimization algorithm tends not to stop at interim points, leading to overly relaxed toxicity boundaries.  By sharing the same $\gamma$ with an attenuation factor of 3, the design avoids this issue.  Second, using the same $\gamma$ enhances optimization efficiency due to the smaller parameter space for optimization. Speed is critical for practical use and the development of web-based software.


\subsection{Optimizing design parameters}\label{sec:optpara}
We choose design parameters $Q = (\lambda_E, \lambda_T, \gamma)$ to maximize power $\beta$ under three type I error constraints specified in equation (\ref{eq:constraint}). In what follows, we first derive a closed-form expression of $\alpha_{00}$, $\alpha_{10}$, $\alpha_{01}$ and $\beta$, and then study the relationship between them and the correlation between $Y_E$ and $Y_T$. These results lead to faster and robust optimization, offering new contributions beyond the original BOP2 paper. 


Suppose at $r$th stage of the trial, we observe $x_{E,r}$ responses and $x_{T,r}$ toxicities from $m_r$ newly treated patients, where $m_r= n_r - n_{r-1}$. Define $\pi_{ET}=Pr(Y_E=1, Y_T=1)$ and conditional probabilities $\pi_{E|T}=Pr(Y_E=1|Y_T=1)$ and $\pi_{E|\bar{T}}=Pr(Y_E=1|Y_T=0)$, which are related as
$\pi_{E|T}=\pi_{ET}/\pi_T$ and $\pi_{E|\bar{T}} = (\pi_E - \pi_{ET})/(1- \pi_T)$. Then the joint probability of $(x_{E,r}, x_{T,r})$ is given as:
\begin{align*}
   d(x_{E,r}, x_{T,r}, m_r, \bm{\pi}) &=b(x_{T,r}, m_r, \pi_T)\\ &\times \sum^{min(x_{E,r},x_{T,r})}_{x=0} b(x, x_{T,r},\pi_{E|T})b(x_{E,r}-x, m_r-x_{T,r}, \pi_{E|\bar{T}}),
\end{align*}
where $\bm{\pi}=(\pi_E, \pi_T, \pi_{ET})$, and $b(\cdot, \cdot, \cdot)$ is the binomial density function. At stage $r$, the go/no-go decision is made by comparing the cumulative number of responses $s_{E,r}$ and the cumulative number of toxicities $s_{T,r}$ to their respective stopping boundaries $(l_{E,r}, l_{T,r})$, which can be predetermined given $Q$. 
Conditional on the trial not stopping at stage $r-1$, ($r>1$, $l_{E,r-1}< s_{E,r-1}\le n_{r-1}$, and $s_{T,r-1} < l_{T,r-1}$), the joint probability of $(s_{E,r}, s_{T,r})$ is 
\begin{align*}
       D(s_{E,r}, s_{T,r},\bm{\pi}) &=\sum_{l_{E,r-1} +1}^{n_{r-1}} \sum_{0}^{l_{T,r-1}-1} d(s_{E,r} -s_{E,r-1},s_{T,r} -s_{T,r-1},n_r-n_{r-1},\bm{\pi} )D(s_{E,r-1}, s_{T,r-1},\bm{\pi}),
\end{align*}
and
\begin{align*}
       D(s_{E,1}, s_{T,1},\bm{\pi}) &=d(s_{E,1},s_{T,1} ,n_{1},\bm{\pi} ).
\end{align*}
Therefore,
\begin{align}
       \alpha_{ij}(Q, H_{ij}, \bm{\pi}) &=Pr\{\text{claim promising} |Q, H_{ij},\bm{\pi} \} \notag\\
        &= \sum_{l_{E,R}+1}^{N} \sum_{0}^{l_{T,R}-1} D(s_{E,R}, s_{T,R},\bm{\pi}), \label{alpha}
\end{align}
where $i =0,1$, $j=0,1$. And 
$$
\beta(Q, H_{11}, \bm{\pi}) =\alpha_{11}(Q, H_{11}, \bm{\pi}).
$$
Given the formula above, the optimal value of $Q$ that maximizes power $\beta$ while satisfying three type I error constraints, as specified in equation (\ref{eq:constraint}), can be quickly determined through grid search. The evaluation of the accuracy of the closed-form expression for Type I errors and power, in comparison with the Monte Carlo approach used in the original BOP2 paper, is provided in Appendix A.3.  More details about the grid search are provided in Appendix A.4.

The determination of the optimal value of $Q$ requires specifying the values of $\pi_E$, $\pi_T$, and $\pi_{ET}$. This is equivalent to specifying $\pi_E$, $\pi_T$, and $\phi$, where $$\phi= \frac{Pr(Y_E =1, Y_T=1)Pr(Y_E=0,Y_T=0)}{Pr(Y_E =1, Y_T=0)Pr(Y_E=0,Y_T=1)}$$ is the odds ratio, representing the correlation between $Y_E$ and $Y_T$. Given $\pi_E$ and $\pi_T$, there exists a one-to-one mapping between $\phi$ and $\pi_{ET}$, with $\phi = \pi_{ET}(1 - \pi_E - \pi_T + \pi_{ET}) / (\pi_E - \pi_{ET})(\pi_T - \pi_{ET})$. Details are provided in Appendix A.5.

Unlike $\pi_E$ and $\pi_T$, whose values are determined by the null and alternative hypotheses (\ref{H00}) and (\ref{H11}), specifying $\phi$ (or equivalently $\pi_{ET}$) can be challenging because the correlation between $Y_E$ and $Y_T$ is often unknown. The following result facilitates the specification of $\phi$.
The proof is provided in Appendix A.6.

\bigskip
\noindent\textbf{Lemma 1.} Let $\bar{\phi}$ denote the true correlation between $Y_T$ and $Y_E$, and assuming that the stopping boundaries of BOP2-TE are determined based on $\phi$, which may be different from $\bar{\phi}$. When $\phi \le \bar{\phi}$,  the type I errors of BOP2-TE are controlled at or below the nominal levels, i.e., $\alpha_{00} \le  \alpha_{00}^*$,  $\alpha_{01} \le  \alpha_{01}^*$, and  $\alpha_{10} \le  \alpha_{10}^*$. 
\bigskip

Therefore, when there is no reliable prior information on the value of $\phi$, we can choose a conservatively smaller value to ensure the type I error control. For example, if $Y_T$ and $Y_E$ are expected to be positively correlated (i.e., $\phi>1$), using $\phi=1$ (i.e., $Y_T$ and $Y_E$ are independent) is a good option guarantee type I error control. Furthermore, our simulation described in Section \ref{simu_study} show that BOP2-TE is remarkably robust: even if $Y_T$ and $Y_E$ are actually negatively correlated (i.e., $\phi<1$), the type I error inflation is very minor or still below the nominal values. Therefore, we recommend $\phi=1$ for general use and set it as the default value in our software. In some cases, clinicians prefer specifying $\pi_{ET}$ rather than $\phi$, due to the former's intuitive interpretation. 
In our software, we allow users to choose one of them as the input.  

\subsection{Software} \label{sec:soft}
We incorporate BOP2-TE into the BOP2 software (see Appendix A.7), available at www.trialdesign.org. The steps of using the software to design a trial are provided as follows:
\begin{enumerate}
    \item Select the Endpoints as ``Efficacy $\&$ Toxicity", and specify the design parameters, including the interim sample size, null values, alternative values, type I error constraints under the \textbf{Trial Setting} tab. 
    \item Obtain the stopping boundaries by clicking the \textbf{Calculate Stopping Boundaries} button.
    \item Generate the operating characteristics of the design by simulation under \textbf{Operating Characteristics} tab.
    \item Go to \textbf{Protocol} tab to retrieve a protocol template that to facilitate the preparation of the trial protocol. 
\end{enumerate}

\section{Application to doses optimization}
BOP2-TE provides a simple and useful design for multiple-dose randomized trials, wherein patients are randomized  to two or three dosages (e.g., the maximum tolerated dose (MTD) and a lower dose) to assess the safety and efficacy of the dosages and determine the optimal dose. Multiple-dose randomized trials has been recommended by FDA's guidance on dose optimization. One challenge of using multiple-dose randomized trials is that it demands a larger sample size, compared to conventional non-randomized dose-finding trials. To mitigate this challenge, the FDA's guidance suggests that ``the use of an adaptive design to stop enrollment of patients to one or more dosage arms of a clinical trial following an interim assessment of efficacy and/or safety could be considered." BOP2-TE offers a streamlined method to facilitate the implementation of this adaptive approach. 

The steps of using BOP2-TE for multiple-dose randomized trials are outlined as follows:
\begin{enumerate}
    \item Specify the design parameters, including the interim time, null and alternative values of toxicity and efficacy, and target type I errors, and use the BOP2-TE software to obtain stopping boundaries;
    \item Randomize patients to dosage arms, and perform interim monitoring and stop the dosage arms that cross the stopping boundaries;
    \item At the end of the trial, select the optimal dose, based on a certain risk-benefit criteria, from the doses that do not cross the stopping boundaries. 
\end{enumerate}

In Step 2, we stop dosage arms early by comparing them with the fixed toxicity and efficacy rates specified in Step 1. We do not consider early stopping by comparing efficacy and toxicity between dosage arms (e.g., whether the high dose arm is significantly more efficacious than the low dose arm). This is because, given a typical maximum sample size of 20-30 per dosage arm, there is little power at the interims (e.g., with 10 patients per arm) to establish that one dosage is better than another for drawing early conclusions and stopping. 

In Step 3, assuming that efficacy $\pi_E(d)$ is nondecreasing with the dosage $d$, we consider the following criterion to select the optimal dose: 
\begin{equation}
d_{opt} = \arg\min_{d} (\hat{\pi}_{E}{(d)}  > \delta \hat{\pi}_{E}{(d_h)}), \label{tradeoff}
\end{equation}
where $d_h$ is the highest dosage under study, which typically is the MTD,  and $\delta$ represents the equivalence margin. Under the assumption that $\pi_E(d)$ is nondecreasing, $d_{opt}$ represents the dosage where the efficacy approximately plateaus. Following the FDA's guidance for bioequivalence studies, we recommend $\delta = 0.8$ as the default value. For a particular trial, it is important to fine-tune the value of $\delta$ through simulation to align with the trial's specific characteristics and objectives. 

It's worth noting that there is no one-size-fits-all risk-benefit criterion for determining the optimal dose; alternative criteria can be considered. For instance, the utility offers a versatile approach to defining the toxicity-efficacy tradeoff and optimal dose \citep{Zhou2019, Lin2020}, generally applicable regardless whether $\pi_E(d)$ is nondecreasing or not.

Due to the randomness of observations, the dose-toxicity and dose-response monotonicity assumption may be violated during interim analyses in step 2 or at the end of trial in step 3. In such situations, we perform isotonic regression on the posterior mean $(\hat{\pi}_E(d), \hat{\pi}_T(d))$ using the pool-adjacent-violators algorithm \citep{Barlow1973} and denote the transformed values as $ \{\tilde{\pi}_T(d),\tilde{\pi}_E(d) \}$. We then compare the $ \{n\tilde{\pi}_T(d),n\tilde{\pi}_E(d) \}$ to the corresponding boundaries to make go/no-go decisions in step 2, where $n$ is the interim sample size. In step 3, we simply replace $\hat{\pi}_E(d)$ with $\tilde{\pi}_E(d)$ to select the optimal dose.
 
There is a valid concern that relying solely on a single risk-benefit criterion like (\ref{tradeoff}) may not be comprehensive enough to encompass all the factors that play into the selection of the optimal dose. In practice, determining the optimal dose is a multifaceted process, encompassing statistical and clinical aspects, as well as both quantitative and qualitative considerations. These considerations encompass a range of endpoints, including toxicity, efficacy, pharmacokinetics (PK), pharmacodynamics (PD), and tolerability. Nonetheless, we hold the view that, even though it may appear somewhat limiting, establishing a risk-benefit criterion for optimal dose selection remains valuable. This is because it allows investigators to conduct simulations to evaluate the performance of the design and make necessary adjustments to design parameters and decision rules. By going through this process, there's a greater likelihood of arriving at a design with appropriate performance characteristics, rather than leaving the optimal dose undefined. The ultimate choice of the optimal dose can then be made based on the design recommendations and additional clinical considerations.

The aforementioned multiple-dose randomized BOP2-TE trial design can be seamlessly integrated with phase I trials to enhance the efficiency of dose optimization. To illustrate, an initial step could involve the application of the Bayesian optimal interval (BOIN, \cite{BOIN2015}) design for dose escalation and the identification of the MTD. Subsequently, the process seamlessly transitions to multiple-dose randomization trials utilizing BOP2-TE, wherein patients are randomized to receive either the MTD or a lower dose. We have conducted a brief evaluation of the operational performance of this integrated BOIN-BOP2-TE dose optimization design in Section \ref{sec:simu:do}.

\section{Operating Characteristics}
\label{simu_study}
\subsection{Jointly monitoring toxicity and efficacy}

We assessed the performance of BOP2-TE across eight scenarios featuring varying null and target response rates as well as toxicity rates (Table \ref{tb:boundary}). The total sample size for these evaluations was set at 36. An interim futility analysis was conducted upon enrolling 18 patients, while two interim toxicity analyses were carried out at patient counts of 9 and 18. Our design optimization considered two different sets of type I error constraints: ($\alpha_{00}^*, \alpha_{01}^*, \alpha_{10}^*$) = (0.025, 0.10, 0.10) and (0.025, 0.10, 0.20).

We evaluated the performance of BOP2-TE using three metrics: 1) probability of claiming the drug is promising (PCP), representing the power under $H_{11}$ and the type I error rate under $H_{00}$, $H_{01}$, or $H_{10}$; 2) probability of early trial termination (PET); 3) expected sample size (ESS).
PCP, PET and ESS can be analytically calculated after the design parameters of BOP2-TE are determined (see Appendix A.8).

%
Table \ref{tb:boundary} presents the stopping boundaries of BOP2-TE and BOP2, assuming $\phi=1$. In general, BOP2-TE yields more stringent stopping boundaries that demand a higher number of responses or/and a lower number of toxicities to make a ``go" decision. For example, given ($\alpha_{00}^*, \alpha_{01}^*, \alpha_{10}^*)$ = (0.025, 0.10, 0.10), in scenario 1 with $(\eta_E, \eta_E^*, \eta_T, \eta_T^*)$ = $(0.50, 0.20, 0.30, 0.10)$, BOP2-TE requires one additional response and one fewer toxicity to declare the agent promising at the end of the trial. In scenario 2, BOP2-TE permits one less response during the interim, but requires one more and also two fewer toxicity events at the end of the trial. The reason that BOP2-TE permits one less response during the interim is that the response boundary of (3, 10) yields the maximum power of 84 \%. Although the response boundary of (4, 10) would also effectively control the type I error rates, its power is slightly lower (i.e.,  82 \%). 
 
When ($\alpha_{00}^*, \alpha_{01}^*, \alpha_{10}^*)$ = (0.025, 0.10, 0.20), implying a greater tolerance for toxicity, the toxicity boundary in BOP2-TE becomes less strict, while the response boundary remains largely unchanged. For example, in scenarios 4 and 5, 
BOP2-TE allows two additional occurrences of toxicities in the end of the trial under $\alpha_{10}^* =0.20$ compared to $\alpha_{10}^* =0.10$, while the response boundary is the same. 

Table \ref{tb:oc} shows the operating characteristics of BOP2-TE and BOP2. As expected, BOP2 only controlled the global type I error $\alpha_{00}$, under which $\alpha_{01}$ and $\alpha_{10}$ can be undesirably high. In contrast, BOP2-TE simultaneously controlled all three type I errors at or below the nominal values. For example, under the target error rates $(\alpha_{00}^*, \alpha_{01}^*, \alpha_{10}^*)$ =  (0.025, 0.10, 0.10),  in scenario 4, BOP2 controlled $\alpha_{00} \le 0.025$, but $\alpha_{10} =0.17$, which are substantially larger than the nominal level of 0.10. In contrast, BOP2-TE controlled $\alpha_{00} \le 0.025$, $\alpha_{10} \le 0.10$ and $\alpha_{01} \le 0.10$. 
Of note, in order to satisfy three type I error constraints simultaneously, the stopping boundaries of BOP2-TE is dictated by the most strict one.  
Based on Theorem 1,  in most cases considered here, meeting the constraints for $\alpha_{01}$ and $\alpha_{10}$ makes $\alpha_{00}$ well below its target value $\alpha_{00}^*$.

Due to additional type I error constraints,  BOP2-TE has slightly lower power than BOP2.
When we relaxed the target value of $\alpha_{10}^*$ from 0.10 to 0.20 to slightly ease the toxicity boundaries, the power of BOP2-TE improved (ranging from 1\% to 7\%) and became comparable to BOP2 across all scenarios. The extent of changes in the operating characteristics depends on the magnitude of the change in the target type I error and also on other factors such as sample size, frequency of interim analyses, and the values of $(\eta_E, \eta_E^*, \eta_T, \eta_T^*)$. This can be easily evaluated using the software provided.

\begin{figure}[H]
  \centering
    \includegraphics[width=1.0\textwidth]{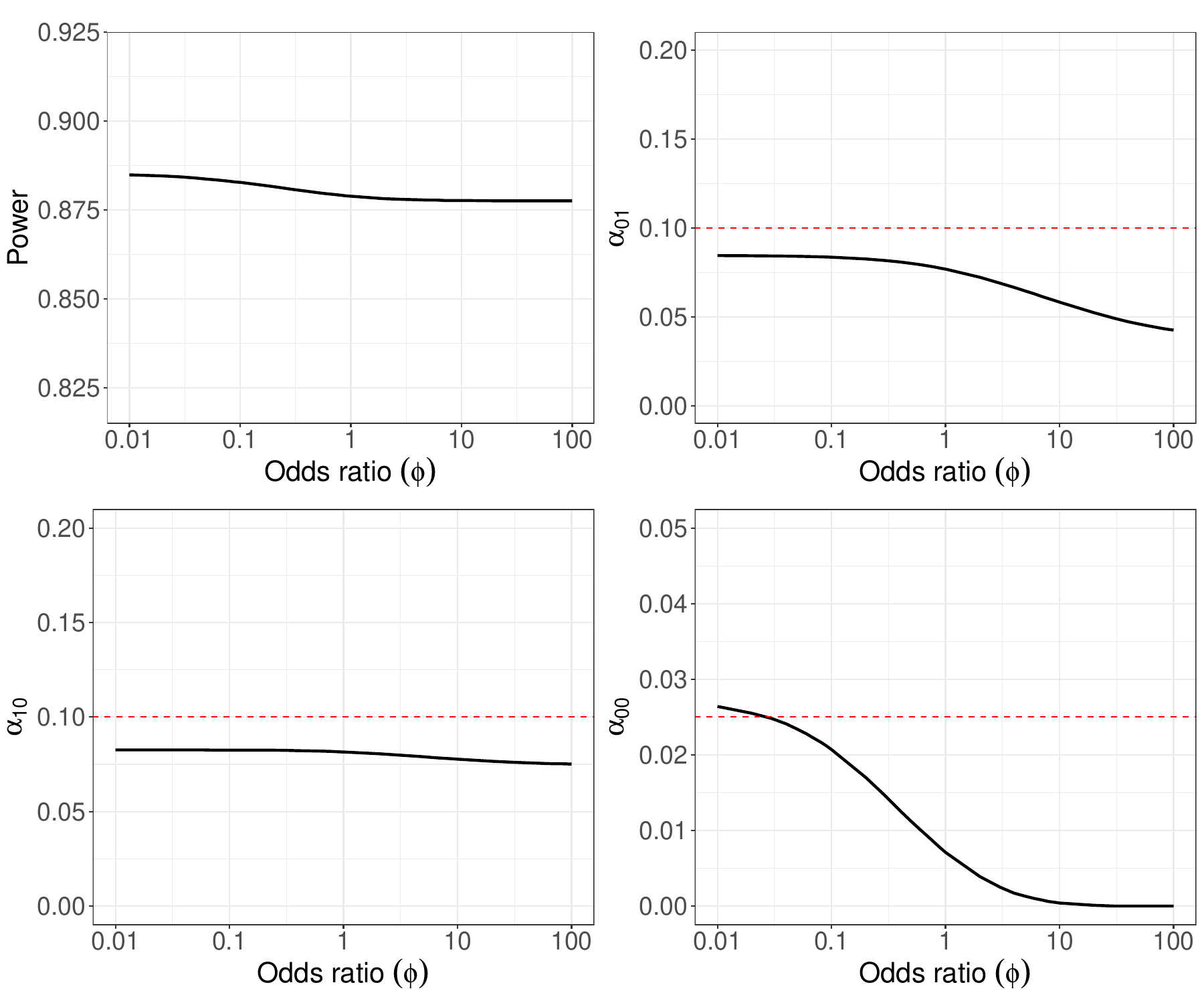}
    \caption{ Sensitivity analysis of power and type I errors when the correlation between efficacy and toxicity (i.e., $\phi$) is misspecified. The horizontal dashed lines indicate the nominal values of type I errors. The design is optimized under the independent assumption with $\phi=1$ and type I error constraints $\alpha_{00}^*=0.025$ and $\alpha_{01}^* = \alpha_{10}^*=0.10$.}
    \label{fg:sensitivity}
\end{figure}

A sensitivity analysis was conducted to evaluate the robustness of BOP2-TE to the misspecification of $\phi$ (i.e., the correlation between $Y_T$ and $Y_E$).  Figure \ref{fg:sensitivity} displays the power and type I error rates under different values of $\phi$ when the parameters of BOP2-TE were determined based on $\phi=1$ (i.e., $Y_T$ and $Y_E$ are independent) under scenario 4 with type I error constraints $\alpha_{00}^*=0.025$ and $\alpha_{01}^* = \alpha_{10}^*=0.10$.  One interim analysis was conducted upon enrolling 18 patients. The optimized stopping boundaies for efficacy and toxicity are (5,14) and (7,11), respectively, at the interim and the end of the trial. When the true value of $\phi$ is greater than 1 (i.e., $Y_T$ and $Y_E$ are positively correlated), BOP2-TE maintains control of $(\alpha_{00}$, $\alpha_{01}$, $\alpha_{10})$ below their nominal values. 
$\alpha_{00}$ appears to be more sensitive to the correlation $\phi$ than $\alpha_{01}$ and $\alpha_{10}$. This is because, under the global null hypothesis $H_{00}$, characterized by a low response rate and a high toxicity rate, increasing the correlation between efficacy and toxicity causes the probability of (efficacy, no toxicity) to drop more rapidly compared to that under the two local null hypotheses. Consequently, the probability of making a false go decision under $H_{00}$ (i.e., $\alpha_{00}$) decreases more quickly. When the correlation is large (e.g., $\phi > 10$), the occurrence of (efficacy, no toxicity) becomes extremely rare, causing $\alpha_{00}$ to approach zero.

In cases where the true value of $\phi$ is less than 1 (i.e., indicating a negative correlation between $Y_T$ and $Y_E$), there is the possibility of slight inflation in type I errors. However, this inflation is typically minor. Importantly, the power of BOP2-TE exhibits robustness to potential misspecification of $\phi$, with differences amounting to less than 0.01. Therefore, when reliable prior information for specifying $\phi$ is unavailable, we recommend setting $\phi=1$ as a prudent choice.

\subsection{Dose optimization} \label{sec:simu:do}
We evaluated the operating characteristics of multiple-dose randomized BOP2-TE design in six scenarios. In scenarios 1-4, we considered two dosage: $d_H$ (a high dose) and $d_L$ (a low dose). For scenarios 5-6, we expanded our evaluation to three dosage: $d_H$ (a high dose), $d_{L1}$ and $d_{L2}$ (two low doses). The sample size for each dose arm was $n=24$, and one interim futility analyses was conducted upon enrolling 12 patients. The null and alternative values were set as $(\eta_E, \eta_E^*, \eta_T^*, \eta_T) = (0.56, 0.24, 0.42, 0.18)$. Type I errors were set as $(\alpha_{00}^*, \alpha_{01}^*, \alpha_{10}^*) = (0.025, 0.10, 0.10)$.

Table \ref{tb:multipledose} shows the results, demonstrating that BOP2-TE efficiently identifies the optimal dose and allows for early termination of futile or toxic doses. For instance, in scenario 1, where $d_H$ was the optimal dose and $d_L$ was futile, the selection for the optimal dose was 73\%. Additionally, the design had a high likelihood of early termination for the $d_L$ arm, with 48.3\% early stopping. In scenario $2$, both doses exhibited high efficacy rates, but $d_H$ was toxic. the selection of $d_L$ was 85.0\%, and the percent of early stopping for $d_H$ was 41.3\%. In scenario $3$, both doses showed promise in terms of efficacy and toxicity. The selection for the two combined doses was almost 100\%. In scenario 4, neither dose was promising, and BOP2-TE terminated the two arms with high percentage. In scenarios 5 and 6, which have three dosage arms, BOP2-TE still exhibits remarkable performance by selecting the optimal dose while early stopping the futile or toxic doses.

We also evaluate the operating characteristics of BOIN-BOP2-TE dose optimization design in six scenarios. The total sample size  was $64$, with 24 patients enrolled in stage I. In stage II, two dose arms were considered with each having 20 patients, and one interim analyses was conducted when $10$ patients treated. The upper toxicity rate was $\psi_T=0.30$. In stage I, we applied BOIN design to identify the MTD from the five doses. In stage II using BOP2-TE, the null and alternative values were set as$(\eta_E, \eta_E^*, \eta_T^*, \eta_T) = (0.42, 0.18, 1.4\psi_T, 0.6\psi_T)$. Type I errors were set as $(\alpha_{00}^*, \alpha_{01}^*, \alpha_{10}^*) = (0.025, 0.15, 0.15)$.

Table \ref{tb: BOIN-BOP2-TE} presents the operating characteristics of the BOIN-BOP2-TE dose optimization design, as well as those of U-BOIN \citep{Zhou2019}, another two-stage design that combines dose escalation with multiple-dose randomization. The objective of including the latter is to facilitate understanding of the operating characteristics of BOIN-BOP2-TE, not to determine which design is superior. This is because the two designs use different criteria to define the optimal dose: BOIN-BOP2-TE uses (\ref{tradeoff}), while U-BOIN uses a utility (see Appendix A.9), making a direct head-to-head comparison not very meaningful.

In general, BOIN-BOP2-TE demonstrated robust performance in identifying the optimal dose, with the selection ranging from 61.80\% to 78.30\% in scenarios 1 to 5. Scenario 6 represents a null case where none of the doses are both safe and efficacious, and BOIN-BOP2-TE successfully allows for early termination of the trial. The two designs showed similar performance in some scenarios (e.g., scenarios 1 and 2), but notable differences in others due to their use of different criteria for defining the optimal dose. For example, in scenario 4, the true utility score for $d_4$ was 64, and for $d_5$ it was 60, resulting in a slight difference that led to a relatively lower selection rate for $d_4$ in U-BOIN. In contrast, BOIN-BOP2-TE directly selected the lower dose as the optimal dose if it exhibited comparable efficacy to the higher dose. As a result, $d_4$ was chosen 19\% more frequently in BOIN-BOP2-TE.

\section{Discussion}
\label{discussion}
We propose a Bayesian optimal design to jointly monitor toxicity and efficacy endpoints. The BOP2-TE design addresses a limitation of the original BOP2 design by simultaneously controlling three type I errors, thereby enhancing patient safety and benefit. Within this framework, we can explore the trade-off between efficacy and toxicity by choosing appropriate type I error constraints. With a closed-form expression for type I errors, we can optimize parameters rapidly and precisely. Furthermore, this closed form enables us to investigate the impact of the correlation between toxicity and efficacy on type I errors and design power.

One of the benefits of BOP2-TE is that efficacy and toxicity assessments do not have to occur simultaneously. For instance, if toxicity events are severe or fatal, we can conduct toxicity monitoring more frequently by adding more interim toxicity examinations while keeping the efficacy interim examinations twice or three times as common in practice. BOP2-TE can also be extended to multinomial endpoints. For example, in a phase II trial of CAR-T cell treatment, clinicians prefer to classify the response into three categories: complete remission (CR), partial remission (PR), or stable disease or progressive disease (SD/PD). Alongside binary toxicity, there are six possible outcomes, and the Dirichlet-multinomial model can easily handle these situations by adding two more categories to the model.

The BOP2-TE method can be seamlessly integrated into dose optimization trials. In the context of a randomized multiple-dose optimization trial, simulation results illustrate that BOP2-TE effectively reduces the dose if it indicates futility or unacceptable toxicity. When combined with the BOIN within the framework of a seamless two-stage design aimed at selecting the optimal dose, the resulting BOIN-BOP2-TE design demonstrates desirable operating characteristics for identifying the optimal dose.

One potential limitation of BOP2-TE, which also applies to BOP2, is that the stopping boundary is optimized within the family of power functions. As a result, the power of BOP2-TE is not necessarily the global maximum. In principle, we can perform global optimization by numerical search, as the boundaries are integers and bounded by the sample size. However, the computational burden exponentially increases with the number of interims. More importantly, we note that although global optimization achieves the global maximum power, its solution often is not practically sound, as it strongly encourages no stopping at interims to enhance overall power. This defeats the purpose of interim monitoring.

We compared the global optimal boundaries, based on exhaustive numerical search, with those of BOP2-TE (see Appendix A.10). The global optimal boundaries are overly relaxed: they would not stop the trial unless either no responses are observed in the first 10 patients or more than half of the patients show toxicity during interim analysis. If we impose a practical constraint of making a no-go decision when the observed efficacy rate falls below the null value or the observed toxicity rate exceeds its null value, the resulting modified global optimal boundaries closely align with those derived from BOP2-TE, with a power difference of less than 0.01.

Thus, we do not view optimization within the power function family as a limitation, as long as it is acknowledged. In fact, this approach aligns well with the well-established frequentist approach, where almost all existing alpha-spending boundaries, such as the Lan-Demets boundaries \citep{Lan1983}, Wang-Tsiatis boundaries \citep{Wang1987}, and Kim-Demets boundaries \citep{Kim1987}, take or approximately take the form of the power function family.

Lastly, this paper focuses on categorical endpoints and binary go/no-go decisions. It is of interest to extend BOP2-TE to continuous and time-to-event endpoints and accommodate trinary decisions (i.e., go/consider/no-go).

\newpage

\begin{table}[]
\begin{adjustbox}{width=\columnwidth,center}
\begin{threeparttable}
\centering
\caption{Stopping boundaries of BOP2-TE and BOP2 under eight scenarios. }
\label{tb:boundary}
\begin{tabular}{cclccccccc}
\hline \hline
 &  &  &  & \multicolumn{2}{c}{$\#$ Response $\le$} &  & \multicolumn{3}{c}{$\#$ Toxicity $\ge$} \\ \cline{5-6} \cline{8-10}
{Scenario} & {$\eta_E, \ \eta_E^*, \ \eta_T^*, \ \eta_T$} & {Method} &  & {18} & {36} &  & {9} & {18} & {36} \\ \hline
 1 &  & BOP2 & & 3 & 9 &  &3  & 5 &  9  \\
 & 0.50, 0.20, 0.30, 0.10 & BOP2-TE & & 3 & 10 &  & 3 & 5 & 8   \\
 &  & BOP2-TE$^1$ &  & 3 & 10 &  & 3 & 6 & 9  \\ \\
2 &  & BOP2 &  & 4 & 9 &  & 4 & 7 & 13 \\
 & 0.50, 0.20, 0.40, 0.20 & BOP2-TE & & 3 & 10 &  & 4 & 7 & 11  \\
 &  & BOP2-TE$^1$ &  & 3 & 10 &  & 4 & 8 & 13  \\
 &  &  &  &  &  &  &  &  &  \\
3 &  & BOP2 &  & 4 & 13 &  & 3 & 5 & 9   \\
 & 0.60, 0.30, 0.30, 0.10 & BOP2-TE &  & 5 & 14 &  & 3 & 5 & 8  \\
 &  & BOP2-TE$^1$ &  & 5 & 14 &  & 3 & 6 & 9  \\
 &  &  &  &  &  &  &  &  &  \\
4 &  & BOP2 &  & 5 & 13 &  & 4 & 7 & 13  \\
 & 0.60, 0.30, 0.40, 0.20 & BOP2-TE & & 5 & 14 &  & 4 & 7 & 11 \\
 &  & BOP2-TE$^1$ &  & 5 & 14 &  & 4 & 8 & 13  \\ \\
5 &  & BOP2 & & 6 & 17 &  & 3 & 5 & 9  \\
 & 0.70, 0.40, 0.35, 0.15 & BOP2-TE &  & 6 & 18 &  & 4 & 6 & 9  \\
 &  & BOP2-TE$^1$ &   & 6 & 18 &  & 4 & 7 & 11   \\ \\
6 &  & BOP2 &  & 6 & 17 &  & 4 & 7 & 12   \\
 & 0.70, 0.40, 0.40, 0.20 & BOP2-TE & & 6 & 18 &  & 4 & 7 & 11    \\
 &  & BOP2-TE$^1$ & & 6 & 18 &  & 4 & 8 & 13  \\ \\
7 &  & BOP2 &  & 8 & 21 &  & 4 & 6 & 10  \\
 & 0.80, 0.50, 0.35, 0.15 & BOP2-TE &  & 8 & 22 &  & 4 & 6 & 9  \\
 &  & BOP2-TE$^1$ &  & 8 & 22 &  & 4 & 7 & 11  \\
 &  &  &  &  &  &  &  &  &  \\
8 &  & BOP2 &  & 8 & 21 &  & 4 & 7 & 12  \\
 & 0.80, 0.50, 0.40 , 0.20 & BOP2-TE &  & 8 & 21 &  & 4 & 7 & 11  \\
 &  & BOP2-TE$^1$ &  & 8 & 22 &  & 4 & 8 & 13  \\ \hline \hline
\end{tabular}%
\begin{tablenotes}
 \item Notes: BOP2 is optimized under $\alpha_{00}^* = 0.025$; BOP2-TE is optimized under $(\alpha_{00}^*, \alpha_{01}^*,  \alpha_{10}^*) =  (0.025, 0.10, 0.10)$; BOP2-TE$^1$ represents BOP2-TE optimized under $(\alpha_{00}^*, \alpha_{01}^*,  \alpha_{10}^*) =  (0.025, 0.10, 0.20)$. All designs assume $\phi=1$.
\end{tablenotes}
\end{threeparttable}
\end{adjustbox}
\end{table}

\begin{table}[]
\begin{adjustbox}{width=\columnwidth,center}
\begin{threeparttable}
\caption{Operating characteristics of BOP2-TE and BOP2, including the probability of claiming the drug is promising (PCP), the probability of early termination (PET), and the expected sample size (ESS).} 
\label{tb:oc} 

\begin{tabular}{ccclccccccccccc}
\hline \hline
 &  &  &  & \multicolumn{3}{c}{BOP2} &  & \multicolumn{3}{c}{BOP2-TE} &  & \multicolumn{3}{c}{BOP2-TE$^1$} \\ \cline{5-7} \cline{9-11} \cline{13-15} 
Scenario & Hypothesis & ($\pi_E$, $\pi_T$) &  & PCP & PET & ESS &  & PCP & PET & ESS &  & PCP & PET & ESS \\ \hline 
1 & $H_{00}$ & (0.20, 0.30) &  & 0.02 & 0.87 & 15.6 &  & 0.01 & 0.86 & 15.6 &  & 0.01 & 0.81 & 16.5 \\
 & $H_{01}$ & (0.20, 0.10) &  & 0.14 & 0.54 & 25.8 &  & 0.08 & 0.53 & 25.9 &  & 0.08 & 0.53 & 26.0 \\
 & $H_{10}$ & (0.50, 0.30) &  & 0.13 & 0.73 & 18.0 &  & 0.09 & 0.73 & 18.1 &  & 0.15 & 0.63 & 19.9 \\
 & $H_{11}$ & (0.50, 0.10) &  & 0.93 & 0.07 & 34.3 &  & 0.92 & 0.07 & 34.3 &  & 0.93 & 0.06 & 34.5 \\ \\
2 & $H_{00}$ & (0.20, 0.40) &  & 0.02 & 0.92 & 14.9 &  & 0.01 & 0.85 & 16.1 &  & 0.02 & 0.80 & 16.9 \\
 & $H_{01}$ & (0.20, 0.20) &  & 0.11 & 0.76 & 21.7 &  & 0.07 & 0.55 & 25.2 &  & 0.08 & 0.55 & 25.4 \\
 & $H_{10}$ & (0.50, 0.40) &  & 0.17 & 0.70 & 18.9 &  & 0.07 & 0.70 & 18.8 &  & 0.19 & 0.60 & 20.5 \\
 & $H_{11}$ & (0.50, 0.20) &  & 0.88 & 0.12 & 33.2 &  & 0.84 & 0.11 & 33.2 &  & 0.89 & 0.09 & 33.6 \\ \\
3 & $H_{00}$ & (0.30, 0.30) &  & 0.02 & 0.82 & 16.5 &  & 0.01 & 0.87 & 15.5 &  & 0.01 & 0.83 & 16.3 \\
 & $H_{01}$ & (0.30, 0.10) &  & 0.14 & 0.37 & 28.8 &  & 0.08 & 0.57 & 25.4 &  & 0.08 & 0.56 & 25.5 \\
 & $H_{10}$ & (0.60, 0.30) &  & 0.14 & 0.08 & 0.73 &  & 0.08 & 0.73 & 18.1 &  & 0.15 & 0.63 & 19.9\\
 & $H_{11}$ & (0.60, 0.10) &  & 0.93 & 0.07 & 34.3 &  & 0.91 & 0.07 & 34.2 &  & 0.93 & 0.06 & 34.4\\ \\
4 & $H_{00}$ & (0.30, 0.40) &  & 0.02 & 0.86 & 15.9 &  & 0.01 & 0.86 & 15.9 &  & 0.02 & 0.81 & 16.7 \\
 & $H_{01}$ & (0.30, 0.20) &  & 0.12 & 0.59 & 24.6 &  & 0.07 & 0.58 & 24.7 &  & 0.08 & 0.58 & 24.9 \\
 & $H_{10}$ & (0.60, 0.40) &  & 0.17 & 0.70 & 18.7 &  & 0.07 & 0.70 & 18.8 &  & 0.19 & 0.60 & 20.5 \\
 & $H_{11}$ & (0.60, 0.20) &  & 0.88 & 0.12 & 33.1 &  & 0.83 & 0.11 & 33.2 &  & 0.89 & 0.10 & 33.5 \\ \\
5 & $H_{00}$ & (0.40, 0.35) &  & 0.02 & 0.84 & 17.3 &  & 0.01 & 0.80 & 18.2 &  & 0.02 & 0.71 & 19.7 \\
 & $H_{01}$ & (0.40, 0.15) &  & 0.12 & 0.58 & 25.2 &  & 0.07 & 0.41 & 28.3 &  & 0.08 & 0.40 & 28.5\\
 & $H_{10}$ & (0.70. 0.35) &  & 0.17 & 0.68 & 20.3 &  & 0.06 & 0.67 & 20.3 &  & 0.19 & 0.54 & 22.8 \\
 & $H_{11}$ & (0.70, 0.15) &  & 0.93 & 0.07 & 34.5 &  & 0.88 & 0.06 & 34.6 &  & 0.94 & 0.04 & 35.0 \\ \\
6 & $H_{00}$ & (0.40, 0.40) &  & 0.02 & 0.81 & 16.9 &  & 0.01 & 0.81 & 16.8 &  & 0.02 & 0.75 & 17.8 \\
 & $H_{01}$ & (0.40, 0.20) &  & 0.12 & 0.44 & 27.2 &  & 0.07 & 0.44 & 27.3 &  & 0.07 & 0.43 & 27.5 \\
 & $H_{10}$ & (0.70, 0.40) &  & 0.12 & 0.70 & 18.9 &  & 0.07 & 0.70 & 18.8 &  & 0.19 & 0.60 & 20.5 \\
 & $H_{11}$ & (0.70, 0.20) &  & 0.87 & 0.11 & 33.4 &  & 0.84 & 0.11 & 33.3 &  & 0.89 & 0.09 & 33.6 \\ \\
7 & $H_{00}$ & (0.50, 0.35) &  & 0.01 & 0.81 & 18.0 &  & 0.01 & 0.81 & 17.9 &  & 0.01 & 0.73 & 19.3 \\
 & $H_{01}$ & (0.50, 0.15) &  & 0.11 & 0.44 & 27.7 &  & 0.06 & 0.44 & 27.7 &  & 0.06 & 0.43 & 27.9 \\
 & $H_{10}$ & (0.80, 0.35) &  & 0.12 & 0.50 & 20.4 &  & 0.06 & 0.67 & 20.3 &  & 0.19 & 0.54 & 22.7 \\
 & $H_{11}$ & (0.80, 0.15)  &  & 0.92 & 0.06 & 34.6 &  & 0.88 & 0.06 & 34.6 &  & 0.95 & 0.04 & 34.9 \\ \\
8 & $H_{00}$ & (0.50, 0.40) &  & 0.02 & 0.82 & 16.7 &  & 0.01 & 0.82 & 16.6 &  & 0.01 & 0.76 & 17.6 \\
 & $H_{01}$ & (0.50, 0.20) &  & 0.10 & 0.47 & 26.8 &  & 0.10 & 0.47 & 26.8 &  & 0.06 & 0.46 & 26.9 \\
 & $H_{10}$ & (0.80, 0.40) &  & 0.12 & 0.69 & 19.0 &  & 0.07 & 0.70 & 18.8 &  & 0.19 & 0.60 & 20.5 \\
 & $H_{11}$ & (0.80, 0.20) &  & 0.88 & 0.10 & 33.4 &  & 0.84 & 0.11 & 33.3 &  & 0.89 & 0.09 & 33.6 \\ \hline \hline
\end{tabular}

\begin{tablenotes}
 \item Notes: BOP2 is optimized under $\alpha_{00}^* = 0.025$; BOP2-TE is optimized under $(\alpha_{00}^*, \alpha_{01}^*,  \alpha_{10}^*) =  (0.025, 0.10, 0.10)$; BOP2-TE$^1$ represents BOP2-TE optimized under $(\alpha_{00}^*, \alpha_{01}^*,  \alpha_{10}^*) =  (0.025, 0.10, 0.20)$. Data were generated with $\phi=1$.
\end{tablenotes}
\end{threeparttable}
\end{adjustbox}
\end{table}

\begin{table}[]
\centering
\caption{Simulation results of BOP2-TE in randomized multiple dose optimization trials, including the selection percentage of optimal dose, the early stop percentage per dose, the average number of patients at each dose.}
\label{tb:multipledose}
\begin{tabular}{ccccccc}
\hline \hline
{Scenario} & {Dose} & $(\pi_E,  \pi_T)$ &  & Selection \% & Early stop  \% & \# patients  \\ \hline \\
\multirow{2}{*}{1} & $d_L$ & 0.30, 0.10 &  & 4.5 & 48.3 & 18.1 \\
 & $d_H$ & 0.60, 0.20 &  & 73.0 & 8.0 & 23.1 \\
 &  &  &  &  &  &  \\
\multirow{2}{*}{2} & $d_L$ & 0.60, 0.15 &  & 85.0 & 2.7 & 23.7 \\
 & $d_H$ & 0.65, 0.35 &  & 5.3 & 41.3 & 19.1 \\
 &  &  &  &  &  &  \\
\multirow{2}{*}{3} & $d_L$ & 0.65, 0.10 &  & 81.7 & 0.52 & 23.9 \\
 & $d_H$ & 0.70, 0.20 &  & 17.7 & 7.42 & 23.1\\
 &  &  &  &  &  &  \\
\multirow{2}{*}{4} & $d_L$ & 0.25, 0.15 &  & 0.3 & 71.8 & 15.4 \\
 & $d_H$ & 0.30, 0.20 &  &5.8 & 43.0 & 18.8\\ \\
  \multirow{2}{*}{5} & $d_{L1}$ & 0.25, 0.10 &  & 1.2 & 57.9 & 17.1 \\
 & $d_{L2}$ & 0.55, 0.15 &  &75.0 & 3.1 & 23.6\\ 
 & $d_H$ & 0.60, 0.30 &  &11.2 & 27.0 & 20.1\\ \\
   \multirow{2}{*}{6} & $d_{L1}$ & 0.55, 0.10 &  & 74.6 & 2.9& 23.7 \\
 & $d_{L2}$ & 0.58, 0.25 &  &16.2& 7.6 & 23.1\\ 
 & $d_H$ & 0.60, 0.35 &  &2.8 & 42.4 & 18.9\\ \hline\hline
\end{tabular}%
\end{table}


\begin{table}[]
\centering
\caption{Simulation results of the BOIN-BOP2-TE design, including the selection percentage of optimal dose, the average number of patients at each dose. The bold in each scenario is the optimal dose.}
\label{tb: BOIN-BOP2-TE}
\resizebox{0.8\textwidth}{!}{%
\begin{tabular}{lllccccc} \hline \hline
 & \textbf{Design} &  & $d_1$ & $d_2$ & $d_3$ & $d_4$ & $d_5$ \\ \hline
 &  &  & \multicolumn{5}{c}{\textbf{Scenario 1}} \\
 &  & DLT rate & \textbf{0.20} & 0.30 & 0.45 & 0.45 & 0.50 \\
 &  & Efficacy rate & \textbf{0.55} & 0.58 & 0.60 & 0.65 & 0.70 \\ \hline
 & BOIN-BOP2-TE & Selection  $\%$ & \textbf{61.80} & 17.40 & 1.40 & 0.90 & 0.00 \\
 &  & No. patients & \textbf{25.00} & 20.40 & 6.60 & 1.60 & 0.30 \\
 & U-BOIN & Selection  $\%$ & \textbf{65.50} & 26.90 & 2.40 & 1.70 & 0.40 \\
 &  & No. patients & \textbf{27.60} & 17.90 & 4.60 & 0.90 & 0.20 \\
 &  &  & \multicolumn{5}{c}{\textbf{Scenario 2}} \\
 &  & DLT rate & 0.08 & \textbf{0.15} & 0.30 & 0.45 & 0.50 \\
 &  & Efficacy rate & 0.25 & \textbf{0.55} & 0.56 & 0.60 & 0.70 \\ \hline
 & BOIN-BOP2-TE & Selection  $\%$ & 2.00 & \textbf{63.80} & 17.70 & 1.60 & 0.10 \\
 &  & No. patients & 9.90 & \textbf{22.60} & 20.80 & 6.70 & 1.10 \\
 & U-BOIN & Selection  $\%$ & 2.30 & \textbf{68.20} & 23.90 & 3.50 & 1.80 \\
 &  & No. patients & 7.90 & \textbf{25.50} & 19.90 & 5.30 & 1.00 \\
 &  &  & \multicolumn{5}{c}{\textbf{Scenario 3}} \\
 &  & DLT rate & 0.05 & 0.08 & \textbf{0.15} & 0.30 & 0.50 \\
 &  & Efficacy rate & 0.25 & 0.35 & \textbf{0.56} & 0.60 & 0.65 \\ \hline
 & BOIN-BOP2-TE & Selection  $\%$ & 0.90 & 7.00 & \textbf{64.20} & 17.60 & 0.30 \\
 &  & No. patients & 4.30 & 10.00 & \textbf{22.50} & 19.90 & 5.00 \\
 & U-BOIN & Selection  $\%$ & 3.40 & 10.40 & \textbf{56.70} & 25.20 & 4.30 \\
 &  & No. patients & 5.50 & 9.40 & \textbf{23.30} & 19.00 & 4.30 \\
 &  &  & \multicolumn{5}{c}{\textbf{Scenario 4}} \\
 &  & DLT rate & 0.01 & 0.03 & 0.05 & \textbf{0.10} & 0.25 \\
 &  & Efficacy rate & 0.10 & 0.15 & 0.20 & \textbf{0.47} & 0.50 \\ \hline
 & BOIN-BOP2-TE & Selection  $\%$ & 0.00 & 0.10 & 1.20 & \textbf{78.30} & 18.10 \\
 &  & No. patients & 3.10 & 3.50 & 7.20 & \textbf{24.90} & 24.70 \\
 & U-BOIN & Selection  $\%$ & 0.70 & 1.80 & 2.80 & \textbf{59.30} & 35.40 \\
 &  & No. patients & 3.70 & 4.50 & 5.80 & \textbf{21.10} & 25.30 \\
 &  &  & \multicolumn{5}{c}{\textbf{Scenario 5}} \\
 &  & DLT rate & 0.01 & 0.03 & 0.05 & 0.07 & \textbf{0.15} \\
 &  & Efficacy rate & 0.10 & 0.15 & 0.20 & 0.25 & \textbf{0.50} \\ \hline
 & BOIN-BOP2-TE & Selection  $\%$ & 0.00 & 0.00 & 1.10 & 11.40 & \textbf{77.40} \\
 &  & No. patients & 3.10 & 3.50 & 4.70 & 24.00 & \textbf{28.70} \\
 & U-BOIN & Selection  $\%$ & 0.80 & 1.90 & 3.80 & 5.90 & \textbf{87.40} \\
 &  & No. patients & 3.80 & 4.70 & 5.60 & 9.00 & \textbf{33.30} \\
 &  &  & \multicolumn{5}{c}{\textbf{Scenario 6}} \\
 &  & DLT rate & 0.22 & 0.45 & 0.55 & 0.65 & 0.70 \\
 &  & Efficacy rate & 0.03 & 0.10 & 0.20 & 0.35 & 0.40 \\ \hline
 & BOIN-BOP2-TE & Selection  $\%$ & 0.00 & 0.10 & 0.10 & 0.00 & 0.00 \\
 &  & No. patients & 31.90 & 12.80 & 1.80 & 0.10 & 0.00 \\
 & U-BOIN & Selection  $\%$ & 0.80 & 5.50 & 1.70 & 0.00 & 0.00 \\
 &  & No. patients & 14.30 & 9.70 & 1.60 & 0.10 & 0.10 \\ \hline \hline
\end{tabular}%
}
\end{table}

\newpage

\vspace{10mm}
\begin{center}
\section*{Appendix}
\end{center}
\renewcommand{\thesection}{A}
\renewcommand{\thetable}{A\arabic{table}}
\setcounter{table}{0}

\subsection{Proof of Theorem 1} \label{A: a00}
 When toxicity and efficacy are independent ($\phi=1$), the conditional probabilities $\pi_{E|T}=\pi_{E|\bar{T}}$. Therefore, the joint probability $d(x_{E,r}, x_{T,r},m_r,\bm{\pi})$ simplifies to:
$$
d(x_{E,r}, x_{T,r}, m_r, \bm{\pi})=b(x_{E,r},m_r,\pi_E)b(x_{T,r},m_r,\pi_T).
$$
Given this simplification, the joint probability of $D(s_{E,r}, s_{T,r})$ is:
\begin{align*}
       D(s_{E,r}, s_{T,r},\bm{\pi}) =\sum_{l_{E,r-1} +1}^{n_{r-1}} \sum_{0}^{l_{T,r-1}-1} &b(s_{E,r} -s_{E,r-1},n_r-n_{r-1},\pi_E )b(s_{T,r} -s_{T,r-1},n_r-n_{r-1},\pi_T ) \\
       & \times D(s_{E,r-1}, s_{T,r-1},\bm{\pi}).
\end{align*}
Thus, for $\alpha_{ij}(Q, H_{ij}, \bm{\pi})$, we have:
\begin{align*}
       \alpha_{ij}(Q, H_{ij}, \bm{\pi}) 
        &= \sum_{l_{E,R}+1}^{N} \sum_{0}^{l_{T,R}-1} D(s_{E,R}, s_{T,R},\bm{\pi})\\
        &=\sum_{l_{E,R}+1}^{N} \sum_{0}^{l_{T,R}-1}\sum_{l_{E,R-1} +1}^{n_{R-1}} \sum_{0}^{l_{T,R-1}-1} b(s_{E,R} -s_{E,R-1},n_R-n_{R-1},\pi_{E,i} ) \\ 
        & \times b(s_{T,R} -s_{T,R-1},n_R-n_{R-1},\pi_{T,j} ) D(s_{E,R-1}, s_{T,R-1},\bm{\pi}) \\
        &=\sum_{l_{E,R}+1}^{n_R}\sum_{l_{E,R-1} +1}^{n_{R-1}}...\sum_{l_{E,1}+1}^{n_1} \prod_{r=1}^R b(s_{E,R}-s_{E,R-1},n_R - n_{R-1},\pi_{E,i})\\
        &\times \sum^{l_{T,R}-1}_0\sum^{l_{T,R-1}-1}_0...\sum^{l_{T,1}-1}_{0}\prod_{r=1}^R b(s_{T,R}-s_{T,R-1},n_R - n_{R-1},\pi_{T,j}) \\
        &=\alpha_E(Q,\pi_{E,i})\alpha_T(Q,\pi_{T,j}),
\end{align*}
where $\pi_{E,i}$ and $\pi_{T,j}$ are the corresponding $\pi_E$ and $\pi_T$ under $H_{ij}$, $i=0,1$, $j=0,1$. Here, $\alpha_E$ represents the probability that the drug is declared effective at the trial's conclusion, indicating that $s_{E,r}$ does not exceed the efficacy stopping boundary $l_{E,r}$ for any $r = 1, \ldots, R$. Similarly, $\alpha_T$ represents the probability that the drug is declared safe, indicating that $s_{T,r}$ does not exceed the toxicity stopping boundary $l_{T,r}$.

Then, we can express $\alpha_{00}$ ($Q$, $H_{ij}$ suppressed for brevity) as follows:
\begin{align*}
\alpha_{00} &= \alpha_{E}(\pi_{E,0})\alpha_{T}(\pi_{T,0})\\
            &= \frac{\alpha_{E}(\pi_{E,0})\alpha_{T}(\pi_{T,0})\alpha_{E}(\pi_{E,1})\alpha_{T}(\pi_{T,1})}{\alpha_{E}(\pi_{E,1})\alpha_{T}(\pi_{T,1})}\\
            &= \frac{\alpha_{10}\alpha_{01}}{\alpha_{11}}=\frac{\alpha_{10}\alpha_{01}}{\beta}.
\end{align*} 
where $\beta =\alpha_{11}$ is power.

\newpage
\subsection{The effect of the attenuation factor on safety stopping boundaries} \label{A: AF3}
Table \ref{tb:Afactors} presents the stopping boundaries of BOP2-TE when the attenuation factor ranges from 1 to 4. We considered that the highest acceptable toxicity rate is 0.3, and thus accordingly set $\eta_T = 0.20$ and $\eta_T^* = 0.40$. Combined with different efficacy rates, three scenarios were examined. The toxicity stopping boundaries generated by attenuation factors of 1 or 2 are excessively relaxed, stopping the trial only when 3/3 and 4/6 patients experience toxicity. Attenuation factors of 3 or 4 generate toxicity stopping boundaries that better align with the conventional rule that $2/3$ and $3/6$ toxicities are deemed overly toxic. Therefore, we opt for an attenuation factor of 3 in BOP2-TE.

\begin{table}[]
\centering
\caption{Stopping boundaries under different attenuation factors (AF) for the toxicity monitoring.}
\label{tb:Afactors}
\resizebox{0.95\textwidth}{!}{%
\begin{tabular}{ccccccccccccc}
\hline 
{Scenarios}&  & {Target type I error} &  & \multicolumn{3}{c}{{Response} \textless{}=} &  & \multicolumn{5}{c}{{Toxicity} \textgreater{}=} \\ \cline{1-1} \cline{3-3} \cline{5-7} \cline{9-13} 
$(\bf{\eta_E^*,\eta_E, \eta_T^*, \eta_T})$&  & $(\alpha_{00}^*, \alpha_{01}^*, \bf{\alpha_{10}^*})$ & {AF} & 12 & 24 & 36 &  & 3 & 6 & 12 & 24 & 36 \\ \hline
\multicolumn{1}{c}{\textbf{}} &  &  &  & \textbf{} &  &  &  &  &  &  &  &  \\
\multicolumn{1}{c}{\textbf{}} &  & 0.05, 0.10, 0.10 & 1 & 0 & 2 & 6 &  & 3 & 4 & 6 & 10 & 11 \\
 &  &  & 2 & 0 & 2 & 6 &  & 2 & 3 & 5 & 9 & 12 \\
 &  &  & 3 & 0 & 2 & 6 &  & 2 & 3 & 5 & 9 & 12 \\
 &  &  & 4 & 0 & 2 & 6 &  & 2 & 3 & 5 & 9 & 12 \\ \\
\multicolumn{1}{c}{(0.1, 0.3, 0.4, 0.2)} &  & 0.05, 0.10, 0.15 & 1 & 0 & 2 & 6 &  & 3 & 4 & 6 & 10 & 12 \\
 &  &  & 2 & 0 & 2 & 6 &  & 2 & 4 & 6 & 9 & 12 \\
 &  &  & 3 & 0 & 2 & 6 &  & 2 & 3 & 5 & 9 & 13 \\
 &  &  & 4 & 0 & 2 & 6 &  & 2 & 3 & 5 & 9 & 13 \\ \\
 &  & 0.05, 0.10, 0.20 & 1 & 0 & 2 & 6 &  & 3 & 4 & 6 & 10 & 12 \\
 &  &  & 2 & 0 & 3 & 7 &  & 3 & 4 & 6 & 10 & 13 \\
 &  &  & 3 & 0 & 2 & 6 &  & 2 & 3 & 5 & 9 & 13 \\
 &  &  & 4 & 0 & 2 & 6 &  & 2 & 3 & 5 & 10 & 14 \\ \hline
\multicolumn{1}{c}{\textbf{}} &  &  &  & \textbf{} &  &  &  &  &  &  &  &  \\
\multicolumn{1}{c}{\textbf{}} &  & 0.05, 0.10, 0.10 & 1 & 0 & 2 & 6 &  & 3 & 4 & 6 & 10 & 11 \\
 &  &  & 2 & 0 & 2 & 6 &  & 2 & 3 & 5 & 9 & 12 \\
 &  &  & 3 & 0 & 2 & 6 &  & 2 & 3 & 5 & 9 & 12 \\
 &  &  & 4 & 0 & 2 & 6 &  & 2 & 3 & 5 & 9 & 12 \\ \\
\multicolumn{1}{c}{(0.2, 0.4, 0.4, 0.2)} &  & 0.05, 0.10, 0.15 & 1 & 0 & 2 & 6 &  & 3 & 4 & 6 & 10 & 12 \\
 &  &  & 2 & 0 & 2 & 6 &  & 2 & 4 & 6 & 9 & 12 \\
 &  &  & 3 & 0 & 2 & 6 &  & 2 & 3 & 5 & 9 & 13 \\
 &  &  & 4 & 0 & 2 & 6 &  & 2 & 3 & 5 & 9 & 13 \\ \\
 &  & 0.05, 0.10, 0.20 & 1 & 0 & 2 & 6 &  & 3 & 4 & 6 & 10 & 12 \\
 &  &  & 2 & 0 & 3 & 7 &  & 3 & 4 & 6 & 10 & 13 \\
 &  &  & 3 & 0 & 2 & 6 &  & 2 & 3 & 5 & 9 & 13 \\
 &  &  & 4 & 0 & 2 & 6 &  & 2 & 3 & 5 & 10 & 14 \\ \hline
\multicolumn{1}{c}{\textbf{}} &  &  &  & \textbf{} &  &  &  &  &  &  &  &  \\
\multicolumn{1}{c}{\textbf{}} &  & 0.05, 0.10, 0.10 & 1 & 2 & 7 & 14 &  & 3 & 4 & 6 & 10 & 11 \\
 &  &  & 2 & 2 & 7 & 14 &  & 2 & 3 & 5 & 9 & 12 \\
 &  &  & 3 & 2 & 7 & 14 &  & 2 & 3 & 5 & 9 & 12 \\
 &  &  & 4 & 2 & 7 & 14 &  & 2 & 3 & 5 & 9 & 12 \\ \\
\multicolumn{1}{c}{(0.3, 0.5, 0.4, 0.2)} &  & 0.05, 0.10, 0.15 & 1 & 2 & 7 & 14 &  & 3 & 4 & 6 & 10 & 12 \\
 &  &  & 2 & 2 & 7 & 14 &  & 2 & 4 & 6 & 9 & 13 \\
 &  &  & 3 & 2 & 7 & 14 &  & 2 & 3 & 5 & 9 & 13 \\
 &  &  & 4 & 2 & 7 & 14 &  & 2 & 3 & 5 & 9 & 14 \\ \\
 &  & 0.05, 0.10, 0.20 & 1 & 2 & 7 & 14 &  & 3 & 4 & 7 & 10 & 13 \\
 &  &  & 2 & 2 & 7 & 14 &  & 3 & 4 & 6 & 10 & 13 \\
 &  &  & 3 & 2 & 7 & 14 &  & 2 & 3 & 6 & 10 & 14 \\
 &  &  & 4 & 2 & 7 & 14 &  & 2 & 3 & 6 & 10 & 14 \\ \hline
\end{tabular}
}
\end{table}
\clearpage


\newpage
\subsection{Comparison of the analytic solution of type I errors with Monte Carlo approach} \label{A:MC}

We validate our closed-form analytic solution of type I errors by comparing it with the Monte Carlo method used in the original BOP2.
We consider scenario 4 in Table 1, where $(\eta_E, \eta_E^*, \eta_T, \eta_T^*) = (0.60, 0.30, 0.20, 0.40)$ and $(\alpha_{00}^*, \alpha_{01}^*, \alpha_{10}^*) = (0.025, 0.10, 0.10)$. The total sample size was set at 36, with interim futility analysis at 18 and two interim toxicity analyses at 9 and 18. 
BOP2-TE determined the efficacy stopping boundaries as (5, 14) and the toxicity stopping boundaries as (4, 7, 11). 

Given these boundaries, we calculated operating characteristics (e.g., type I errors, power) using our closed-form analytic formula and the Monte Carlo method. Table \ref{tb:validation} presents the results. When the number of Monte Carlo simulations reached 200,000, the two methods produced nearly identical results in terms of the probability of claiming the drug is promising (PCP, which is type I error or power), the probability of early termination (PET), and the expected sample size (ESS).


\begin{table}[]
\centering
\caption{{Comparison of the proposed closed-form analytic solution and Monte Carlo method (used in the original BOP2) for calculating the probability of claiming the drug is promising (PCP), the probability of early termination (PET), and the expected sample size (ESS). PCP represents type I error under $H_{00}$, $H_{01}$ and $H_{11}$ and power under $H_{11}$. }}
\label{tb:validation}
\begin{tabular}{cccccc} \hline \hline
Hypothesis:$(\pi_E, \pi_T)$ & \multicolumn{1}{l}{Methods} & No. Simulation & \multicolumn{1}{c}{PCP} & \multicolumn{1}{c}{PET} & \multicolumn{1}{c}{ESS} \\ 
 &  & ($\times10^4$) & &  &  \\
 \hline
\multirow{6}{*}{$H_{00}: (0.30,0.40)$}  & \multirow{5}{*}{Monte Carlo} & 1 & 0.0064 & 0.8594 & 15.88 \\
 &  & 5 & 0.0061 & 0.8585 & 15.89 \\
 &  & 10 & 0.0063 & 0.8570 & 15.91 \\
 &  & 15 & 0.0063 & 0.8586 & 15.88 \\
 &  & 20 & 0.0063 & 0.8585 & 15.89 \\
 & \multicolumn{1}{l}{Analytic} & 0 & \textbf{0.0063} & \textbf{0.8586} & \textbf{15.89} \\ \hline
 & \multicolumn{1}{l}{} &  & \multicolumn{1}{c}{} & \multicolumn{1}{c}{} & \multicolumn{1}{c}{} \\
\multirow{6}{*}{$H_{01}: (0.30,0.20)$}  & \multirow{5}{*}{Monte Carlo} & 1 & 0.0777 & 0.5848 & 24.69 \\
 &  & 5 & 0.0748 & 0.5825 & 24.76 \\
 &  & 10 & 0.0741 & 0.5857 & 24.68 \\
 &  & 15 & 0.0731 & 0.5844 & 24.71 \\
 &  & 20 & 0.0730 & 0.5844 & 24.71 \\
 & \multicolumn{1}{l}{Analytic} & 0 & \textbf{0.0728} & \textbf{0.5845} & \textbf{24.71} \\ \hline
 & \multicolumn{1}{l}{} &  & \multicolumn{1}{c}{} & \multicolumn{1}{c}{} & \multicolumn{1}{c}{} \\
\multirow{6}{*}{$H_{10}: (0.60,0.40)$}  & \multirow{5}{*}{Monte Carlo} & 1 & 0.0733 & 0.6966 & 18.79 \\
 &  & 5 & 0.0723 & 0.6997 & 18.75 \\
 &  & 10 & 0.0714 & 0.6982 & 18.77 \\
 &  & 15 & 0.0723 & 0.6976 & 18.78 \\
 &  & 20 & 0.0724 & 0.6978 & 18.78 \\
 & \multicolumn{1}{l}{Analytic} & 0 & \textbf{0.0724} & \textbf{0.6982} & \textbf{18.78} \\ \hline
 & \multicolumn{1}{l}{} &  & \multicolumn{1}{c}{} & \multicolumn{1}{c}{} & \multicolumn{1}{c}{} \\
\multirow{6}{*}{$H_{11}: (0.60,0.20)$}  & \multirow{5}{*}{Monte Carlo} & 1 & 0.8319 & 0.1125 & 33.20 \\
 &  & 5 & 0.8338 & 0.1119 & 33.22 \\
 &  & 10 & 0.8320 & 0.1134 & 33.19 \\
 &  & 15 & 0.8337 & 0.1129 & 33.20 \\
 &  & 20 & 0.8337 & 0.1127 & 33.20 \\
 & \multicolumn{1}{l}{Analytic} & 0 & \textbf{0.8337} & \textbf{0.1127} & \textbf{33.20} \\ \hline \hline
\end{tabular}
\end{table}

\newpage

\subsection{Grid search to determine $\lambda_E$, $\lambda_T$ and $\gamma$}
\label{gridsearch}
In our simulation, the search space for $\lambda_E$ and $\lambda_T$ is $[0.5, 0.99]$. Increments are set at 0.025 from 0.5 to 0.8, and at 0.01 from 0.8 to 0.99. The search space starts from 0.5 because it represents the probability cutoff for $\Pr(\pi_E> \eta_E^*  |D_n)$ and $\Pr(\pi_T \le \eta_T^*  |D_n)$ at the end of the trial. A value of $\lambda_E$ and $\lambda_T$ less than 0.5 represents overly weak evidence (i.e., less than 50\% chance that  $\pi_E> \eta_E^*$ and $\pi_T \le \eta_T^*$) to claim that the treatment is efficacy and safe, leading to unacceptably large type I errors (approximately 0.5 or more).  As a reasonable value is most likely located in the range of 0.8 to 0.99 (approximately corresponding to a Type I error of 0.2 and lower), we use a finer grid in that range.

For the power parameter $\gamma$, the search space extends from $[0, 1]$, with increments determined by $\log(\text{seq}(1, 0.5, \text{by} = -0.025))/\log(0.5)$. This method ensures that $(\frac{n}{N})^{\gamma}$ is distributed more evenly across the search space. Specifically, when $n/N=0.5$, the value of $0.5^{\gamma}$ are even distributed (e.g., 0.5, 0.525,..., 0.975, 1), facilitating consistent increments in the grid search space for $C_E(n)$ and $C_T(n)$. This approach prevents the grid search space from having uneven granularity, reducing the chance of missing the optimal solution.

This configuration results in a total of 17,661 potential parameter combinations, covering practically plausible boundaries. If desired, finer granularity could be explored. The time required for optimization depends on the total sample size and the frequency of interim monitoring. With fewer than six interims and a total sample size of around 50, the search duration varies from 5 to 20 seconds.

\newpage
\subsection{The relationship between $\pi_{ET}$ and $\phi$}
Let $\pi_{11} = \text{Pr}(Y_E =1, Y_T=1)$, $\pi_{10} = \text{Pr}(Y_E =1, Y_T=0)$, $\pi_{01} = \text{Pr}(Y_E =0, Y_T=1)$ and $\pi_{00} = \text{Pr}(Y_E =0, Y_T=0)$. Given the marginal efficacy rate $\pi_E$, marginal toxicity rate $\pi_T$, and joint probability $\pi_{ET}$, as defined in the manuscript, it follows that
\begin{align*}
\pi_{11} &= \pi_{ET} \\
\pi_{10} &= \pi_E - \pi_{ET} \\  
\pi_{01} &= \pi_T - \pi_{ET} \\ 
\pi_{00} &= 1- \pi_E - \pi_T + \pi_{ET}.
\end{align*}
Therefore, the odds ratio $\phi$ is given by:
\begin{align*}
\phi &= \frac{\text{Pr}(Y_E = 1, Y_T = 1)\text{Pr}(Y_E = 0, Y_T = 0)}{\text{Pr}(Y_E = 1, Y_T = 0)\text{Pr}(Y_E = 0, Y_T = 1)} \\
&= \frac{\pi_{11}\pi_{00}}{\pi_{10}\pi_{01}} = \frac{\pi_{ET}(1 - \pi_E - \pi_T + \pi_{ET})}{(\pi_E - \pi_{ET})(\pi_T - \pi_{ET})}.
\end{align*}
This equation provides the transformation from $\pi_{ET}$ to $\phi$.  \\

To transform $\phi$ to $\pi_{ET}$, we re-arrange the above equation as:
$$ \phi((\pi_E - \pi_{ET})(\pi_T - \pi_{ET}))=\pi_{ET}(1 - \pi_E - \pi_T + \pi_{ET}),$$
leading to the quadratic equation:
$$ (1-\phi)\pi_{ET}^2 +(1-(1-\phi)(\pi_E +\pi_T))\pi_{ET} - \phi\pi_E\pi_T.$$
Solving this equation under the constraint $\pi_{ET} > 0$, we obtain
\[
\pi_{ET} = \begin{cases} 
\pi_E \pi_T & \text{if } \phi = 1, \\
\frac{1}{2} \left(\pi_E + \pi_T - \frac{1}{1-\phi} + \sqrt{(\pi_E + \pi_T - \frac{1}{1-\phi})^2 + \frac{4\phi\pi_E \pi_T}{1-\phi}}\right) & \text{if } \phi < 1, \\
\frac{1}{2} \left(\pi_E + \pi_T - \frac{1}{1-\phi} - \sqrt{(\pi_E + \pi_T - \frac{1}{1-\phi})^2 + \frac{4\phi\pi_E \pi_T}{1-\phi}}\right) & \text{if } \phi > 1.
\end{cases}
\]

\newpage
\subsection{Proof of Lemme 1}
Given the one-to-one transformation between the odds ratio $\phi$ and $\pi_{ET}$,  we express $\alpha_{ij}(Q, H_{ij}, \pi_{ET})$ as $\alpha_{ij}(\phi)$, where arguments $Q$ and $H_{ij}$ are suppressed for notational brevity. 

As described in section 2.1, given a Dirichlet prior
$$
    \pi_1, ..., \pi_K \sim \text{Dirichlet}(\tau_1, ...,\tau_K),
$$
the posterior distributions of $\pi_E$ and $\pi_T$ are both beta distributions, as shown in equations (2.1) and (2.2). Given $Q$, stopping boundaries $\{l_{E,1},...,l_{E,R} \}$ and $\{l_{T,1},...,l_{T,R}\}$ can be determined prior to the trial start using numerical search \citep{Zhou2017}. The type I errors then can be calculated using these boundaries as follows:
\begin{align*}
       \alpha_{ij}(\phi)= \sum_{l_{E,R}+1}^{n_R} \sum_{0}^{l_{T,R}-1} D(s_{E,R}, s_{T,R},\bm{\pi}). 
\end{align*}
Given fixed $\{l_{E,1},...,l_{E,R} \}$ and $\{l_{T,1},...,l_{T,R}\}$, \citet{Bryant1993} showed that $\alpha_{ij}(\phi)$ is a non-decreasing function of  $1/\phi$.
 Thus, for any $\phi < \bar{\phi}$
$$
\alpha_{ij}(\bar{\phi}) \le \alpha_{ij}(\phi) \le \alpha_{ij}^{*}.
$$
Therefore, Lemma 1 follows.

\newpage
\subsection{BOP2-TE Shiny application}
\renewcommand{\thefigure}{A\arabic{figure}}
\setcounter{figure}{0}
\begin{figure}[H]
    \hspace{-0.8in}
    \includegraphics[width=1.2\textwidth]{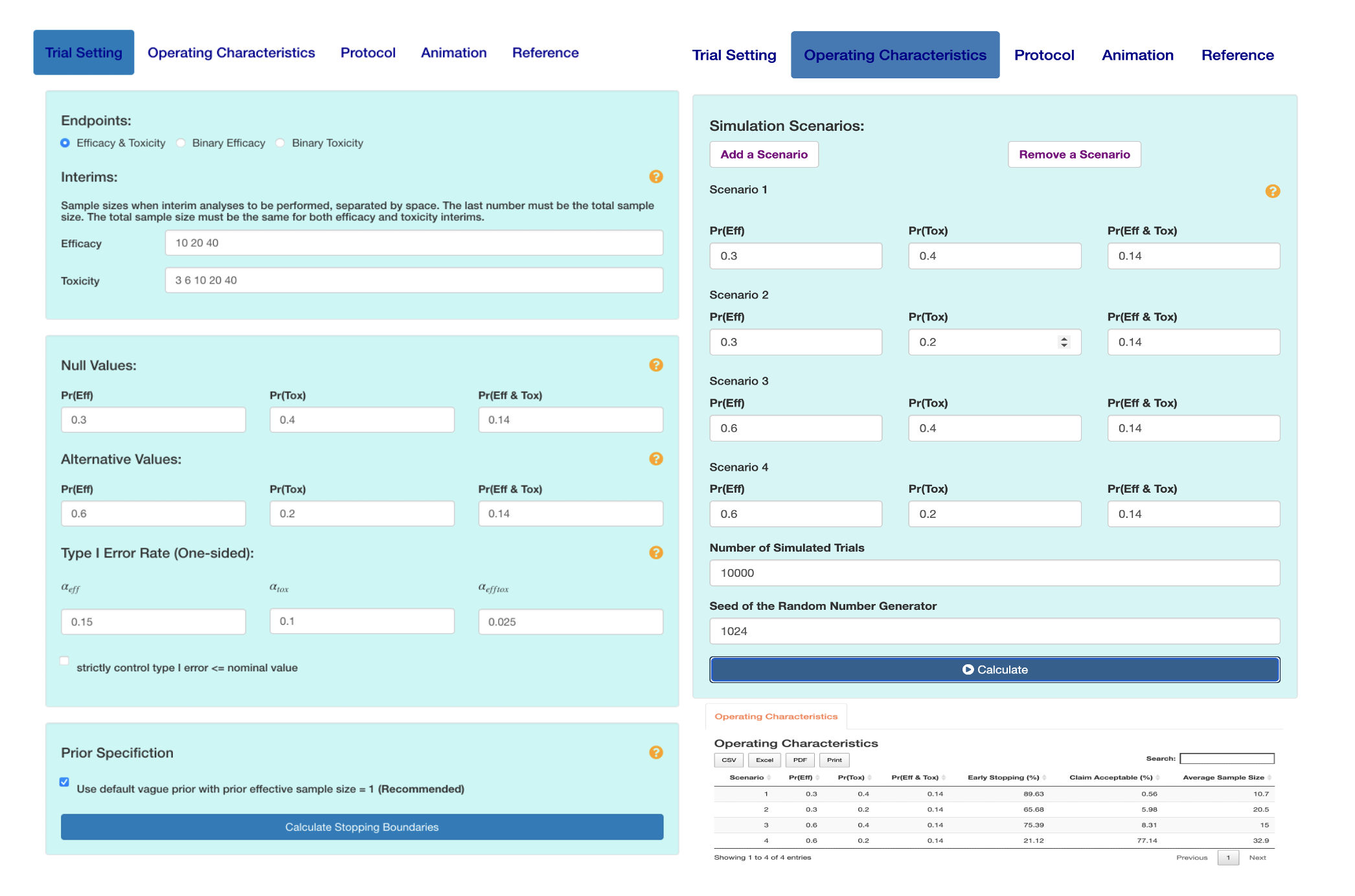}
    \caption{The user interface of BOP2-TE as a module of the BOP2 app.}
    \label{fig:shiny}
\end{figure}

\newpage
\subsection{Analytical calculation of PCP, PET and ESS} \label{A:ESS}
Let $\alpha_{ij(r)}$ represent the probability of claiming promising after stage $r$ in the trial, where $i$, $j$ = 0, 1.   For notational brevity, we suppress the subscripts $i$ and $j$ in the equations presented below. The calculations are given as follows:
\begin{align*}
     &\text{ESS} = m_1 + \sum_{r=2}^R m_r \alpha_{(r-1)}, \\
     &\text{PET} = 1- \alpha_{(R-1)},  \\
     &\text{PCP} = \alpha_R \ .
\end{align*}

\newpage
\subsection{U-BOIN setting} \label{A:U-BOIN}
The simulation study of U-BOIN was conducted using the Shiny app available at \url{https://www.trialdesign.org}. The cohort size was specified as 3, with a total of 21 cohorts. 8 cohorts were used in stage I and 13 cohorts were used in stage II. The upper toxicity rate was $0.30$ and the lower efficacy rate was $0.30$. Utility was specified as follows:
\begin{figure}[H]
    \includegraphics[width=1\textwidth]{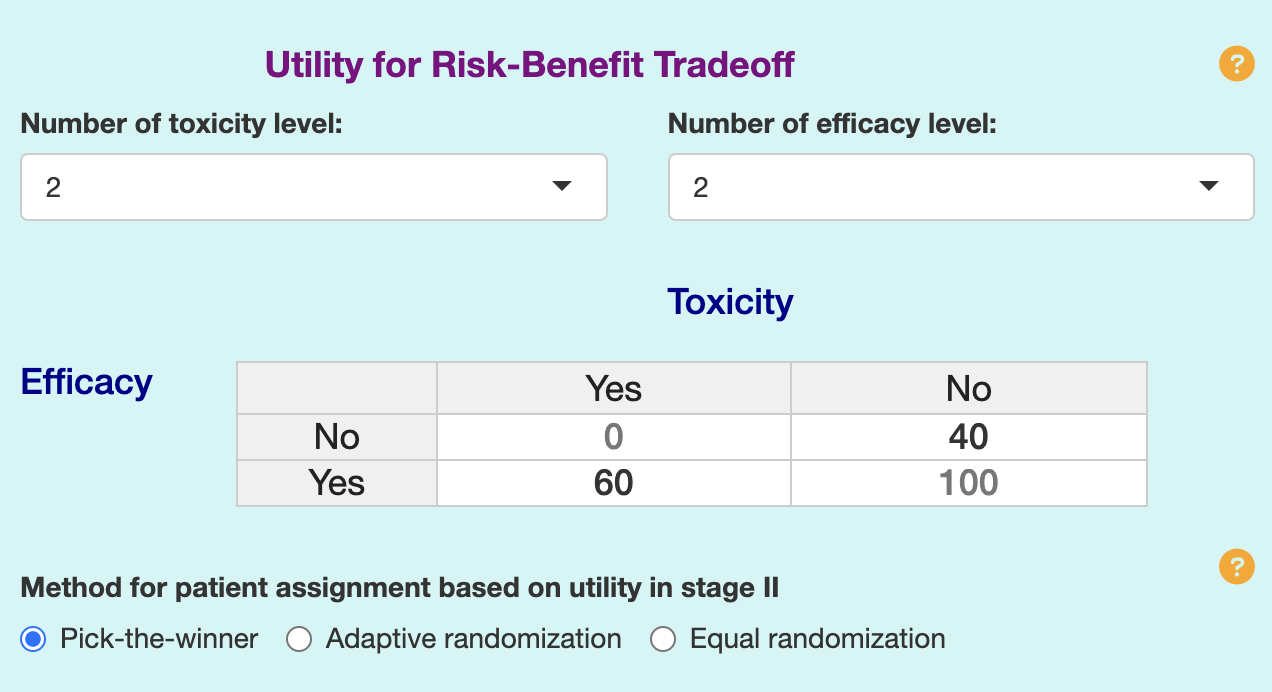}
    \label{fig:U-BOIN}
\end{figure}

\newpage
\subsection{Comparison of the boundaries of BOP2-TE with the globally optimal stopping boundaries}
\begin{table}[h]
\begin{adjustbox}{width=\columnwidth,center}
\begin{threeparttable}
\centering
\caption{The comparison of the boundaries of BOP2-TE with the globally optimal stopping boundaries under scenarios 1-4. }
\label{tb:localoptm}
\begin{tabular}{clccccccccccc} \hline \hline
 Scenario &  & \multicolumn{2}{c}{\# Response $\le$} &  & \multicolumn{3}{c}{\# Toxicity $\ge$} &  &  &  &  &  \\ \cline{3-4} \cline{6-8}
$(\eta_E, \eta_E^*, \eta_T, \eta_T^*)$ &  & 18 & 36 &  & 9 & 18 & 36 &  & $\alpha_{00}$ & $\alpha_{01}$ & $\alpha_{10}$ & power \\ \hline 
 & Global & 0 & 10 &  & 5 & 5 & 8 &  & 0.009 & 0.085 & 0.097 & 0.953 \\
(0.5, 0.2, 0.1, 0.3) & Global* & 3 & 10 &  & 3 & 6 & 8 &  & 0.008 & 0.079 & 0.091 & 0.923 \\
 & BOP2-TE & 3 & 10 &  & 3 & 5 & 8 &  & 0.007 & 0.078 & 0.085 & 0.915 \\
\multicolumn{1}{l}{} &  & \multicolumn{1}{l}{} & \multicolumn{1}{l}{} & \multicolumn{1}{l}{} & \multicolumn{1}{l}{} & \multicolumn{1}{l}{} & \multicolumn{1}{l}{} & \multicolumn{1}{l}{} & \multicolumn{1}{l}{} & \multicolumn{1}{l}{} & \multicolumn{1}{l}{} & \multicolumn{1}{l}{} \\
 & Global & 0 & 10 &  & 7 & 10 & 11 &  & 0.008 & 0.081 & 0.090 & 0.906 \\
(0.5, 0.2, 0.2, 0.4) & Global* & 3 & 10 &  & 4 & 8 & 11 &  & 0.006 & 0.072 & 0.076 & 0.846 \\
 & BOP2-TE & 3 & 10 &  & 4 & 7 & 11 &  & 0.006 & 0.071 & 0.073 & 0.837 \\
\multicolumn{1}{l}{} &  & \multicolumn{1}{l}{} & \multicolumn{1}{l}{} & \multicolumn{1}{l}{} & \multicolumn{1}{l}{} & \multicolumn{1}{l}{} & \multicolumn{1}{l}{} & \multicolumn{1}{l}{} & \multicolumn{1}{l}{} & \multicolumn{1}{l}{} & \multicolumn{1}{l}{} & \multicolumn{1}{l}{} \\
 & Global & 0 & 14 &  & 5 & 5 & 8 &  & 0.009 & 0.088 & 0.097 & 0.950 \\
(0.6, 0.3, 0.1, 0.3) & Global* & 5 & 14 &  & 3 & 6 & 8 &  & 0.008 & 0.080 & 0.091 & 0.919 \\
 & BOP2-TE & 5 & 14 &  & 3 & 5 & 8 &  & 0.007 & 0.080 & 0.085 & 0.912 \\
 &  &  &  &  &  &  &  &  &  &  &  &  \\
 & Global & 0 & 14 &  & 8 & 10 & 11 &  & 0.008 & 0.084 & 0.090 & 0.903 \\
(0.6, 0.3, 0.2, 0.4) & Global* & 5 & 14 &  & 4 & 8 & 11 &  & 0.007 & 0.074 & 0.075 & 0.842 \\
 & BOP2-TE & 5 & 14 &  & 4 & 7 & 11 &  & 0.006 & 0.073 & 0.072 & 0.834 \\ \hline \hline
\end{tabular}
\begin{tablenotes}
 \item Global* represents the global optimal boundaries with the constraints that a no-go decision is made if the observed efficacy rate is less than the null value $\eta_E^*$ or if the observed toxicity rate is larger than the null value $\eta_T^*$.
\end{tablenotes}
\end{threeparttable}
\end{adjustbox}
\end{table}

\end{document}